\documentclass[3p,11pt]{elsarticle}

\usepackage{lineno,hyperref}
\usepackage{xcolor}
\usepackage{lscape}
\modulolinenumbers[5]
\usepackage{color, soul}
\usepackage{rotating}
\usepackage{natbib}
\setcitestyle{authoryear,open={(},close={)}}

\journal{Remote Sensing of Environment}

\DeclareUnicodeCharacter{FB01}{fi}
\begin{document}

\begin{frontmatter}

\title{Satellite data for the offshore renewable energy sector: synergies and innovation opportunities}

\author[mymainaddress]{E. Medina-Lopez}\cortext[mycorrespondingauthor]{Corresponding author}\corref{mycorrespondingauthor}
\ead{emedina@ed.ac.uk}
\author[mysecondaryaddress]{D. McMillan}
\author[mythirdaddress]{J. Lazic}
\author[mysecondaryaddress]{E. Hart}
\author[mymainaddress]{S. Zen}
\author[mymainaddress]{A. Angeloudis}
\author[myfourthaddress]{E. Bannon}
\author[mysecondaryaddress]{J. Browell}
\author[my5address]{S. Dorling}
\author[my6address]{R.M. Dorrell}
\author[my6address]{R. Forster}
\author[my7address]{C. Old}
\author[my8address]{G.S. Payne}
\author[my9address]{G. Porter}
\author[my6address]{A.S. Rabaneda}
\author[my7address]{B. Sellar}
\author[my6address]{E. Tapoglou}
\author[my10address]{N. Trifonova}
\author[my11address]{I.H. Woodhouse}
\author[my10address]{A. Zampollo}

\address[mymainaddress]{
Institute for Infrastructure and Environment, School of Engineering, The University of Edinburgh, The King’s Buildings, Edinburgh, UK}
\address[mysecondaryaddress]{University of Strathclyde}
\address[mythirdaddress]{Bayes Centre, The University of Edinburgh}
\address[myfourthaddress]{Wave Energy Scotland}
\address[my5address]{Weatherquest Ltd.}
\address[my6address]{University of Hull,Cottingham Rd., Hull HU6 7RX UK}
\address[my7address]{Institute for Energy Systems, School of Engineering, The University of Edinburgh, The King’s Buildings, Edinburgh EH9 3JL, UK}
\address[my8address]{Laboratoire de recherche en Hydrodynamique, Energ\'etique et Environnement Atmosph\'erique, Ecole Centrale Nantes, 1 rue de la No\"e, 44300 Nantes, France}
\address[my9address]{Orbital Micro Systems}
\address[my10address]{School of Biological Sciences, University of Aberdeen, Tillydrone Ave, Aberdeen AB24 2TZ, UK}
\address[my11address]{School of Geosciences, The University of Edinburgh, Edinburgh EH9 3JL, UK}

\begin{abstract}
Can satellite data be used to address challenges currently faced by the Offshore Renewable Energy (ORE) sector? What benefit can satellite observations bring to resource assessment and maintenance of ORE farms? Can satellite observations be used to assess the environmental impact of offshore renewables leading towards a more sustainable ORE sector?
This review paper faces these questions presenting a holistic view of the current interactions between satellite and ORE sectors, and future needs to make this partnerships grow. The aim of the work is to start the conversation between these sectors by establishing a common ground. We present offshore needs and satellite technology limitations, as well as potential opportunities and areas of growth. To better understand this, the reader is guided through the history, current developments, challenges and future of offshore wind, tidal and wave energy technologies. Then, an overview on satellite observations for ocean applications is given, covering types of instruments and how they are used to provide different metocean variables, satellite performance, and data processing and integration. Past, present and future satellite missions are also discussed. Finally, the paper focuses on innovation opportunities and the potential of synergies between the ORE and satellite sectors. Specifically, we pay attention to improvements that satellite observations could bring to standard measurement techniques: assessing uncertainty, wind, tidal and wave conditions forecast, as well as environmental monitoring from space. Satellite--enabled measurement of ocean physical processes and applications for fisheries, mammals and birds, and habitat change, are also discussed in depth.
\end{abstract}


\end{frontmatter}


\textnormal{\tableofcontents}
\noindent\rule{\textwidth}{1pt}

\section{Introduction}


The motivation for this study arises from a need for cost reduction in the way metocean data is typically collected in the marine environment (\textit{e.g.} by using offshore meteorological stations, which are expensive to deploy and maintain due to the harsh marine weather conditions), and how this data gathering process affects the total cost of energy in offshore renewable energy projects. Marine data has an impact in all phases of a marine renewable farm development, including initial stages such as site selection and device design, but also construction (particularly during operations and maintenance) and decommissioning. Nowadays, the planning of operations and maintenance (O\&M) missions and vessels dispatching for offshore installations is limited to updated numerical weather forecasts every 6 to 12 hours. Being able to correctly predict, on a short--term, the weather conditions will allow the optimisation of costs and resources for O\&M. Moreover, the correct estimate of access windows that allow the crew to conduct the management operations safely is difficult due to the high degree of uncertainty characterising the modelling tools currently available for managers.\\

Satellite--based measurements have long been identified as having a potential role in enabling cost reduction of marine renewables, but applications have been largely limited to wind resource assessment and wake modelling. This paper aims to take satellite data usage in offshore renewable energy (ORE) to the next level by better linking satellite data, models driven by such data, decisions driven by the model outputs, and quantifying this impact on Levelized Cost of Energy (LCOE). By mapping linkages between key decision horizons in ORE life cycle to satellite capability, this paper presents a map of where satellite data can best impact ORE project decisions. This map will direct the data analysis activities towards the project decisions having the best potential for improvement and quantify any reductions in uncertainty. These improvements can be captured and monetised in a range of cost models.\\

Offshore climate information is derived from various key satellite datasets. Offshore wind resource can be derived from satellite synthetic aperture radar (SAR) for localised use, and satellite scatterometers at a wider scale. SARs can also be used to model wave and tidal resource, and satellite altimeters are specifically used for wave resource characterisation. There is a great range of free and open satellite datasets that are relevant for ORE applications. As an example, Copernicus Sentinel-1 is a SAR that has been operating since 2014, providing coverage over Europe, Canada and main shipping routes in 2--4 days, regardless of weather conditions. SAR data is delivered to Copernicus services within an hour of acquisition, \cite{copernicus}. Its high resolution ($20$m) allows sufficiently detailed coverage of relevant areas. The information can be download from the Sentinel Hub online system and operated locally using the Sentinel Application Platform (SNAP) for satellite data, \cite{snap}, or it can be operated online from Google Earth Engine some days after acquisition, \cite{gee}, reducing operation time. The satellite--derived information can be compared with existing numerical models and \textit{in situ} measurement stations, obtaining a quantitative comparison that is able to feed into life cycle, operations, and cost of energy analyses. \\


The paper contains three sections: the first one presents the ORE sector, describing the offshore wind, tidal and wave sub--sectors. A brief introduction to the history of these fields is provided, followed by current techniques for resource assessment and site characterisation, main technologies and commercial projects, and finally by presenting the challenges of the sector. The second section of the paper covers the satellite observations sector. An overview on available technologies and their use is given, followed by some insight on satellite data processing and integration. The section ends with a summary of the past, present and future of satellite missions and their applicability to metocean observations. The last section of the paper presents the ORE and satellite data synergies, focusing on offshore wind, tidal and wave forecasting, as well as other key parameters that can be improved by the use of satellite data, such as environmental monitoring, physical processes in the marine environment, fisheries, mammals and birds, and habitats. The section ends with a note on uncertainty quantification in the use of satellite data.

\section{The offshore renewable energy industries}
\begin{figure}[!h] 
\centering
\includegraphics[width=0.95\textwidth] {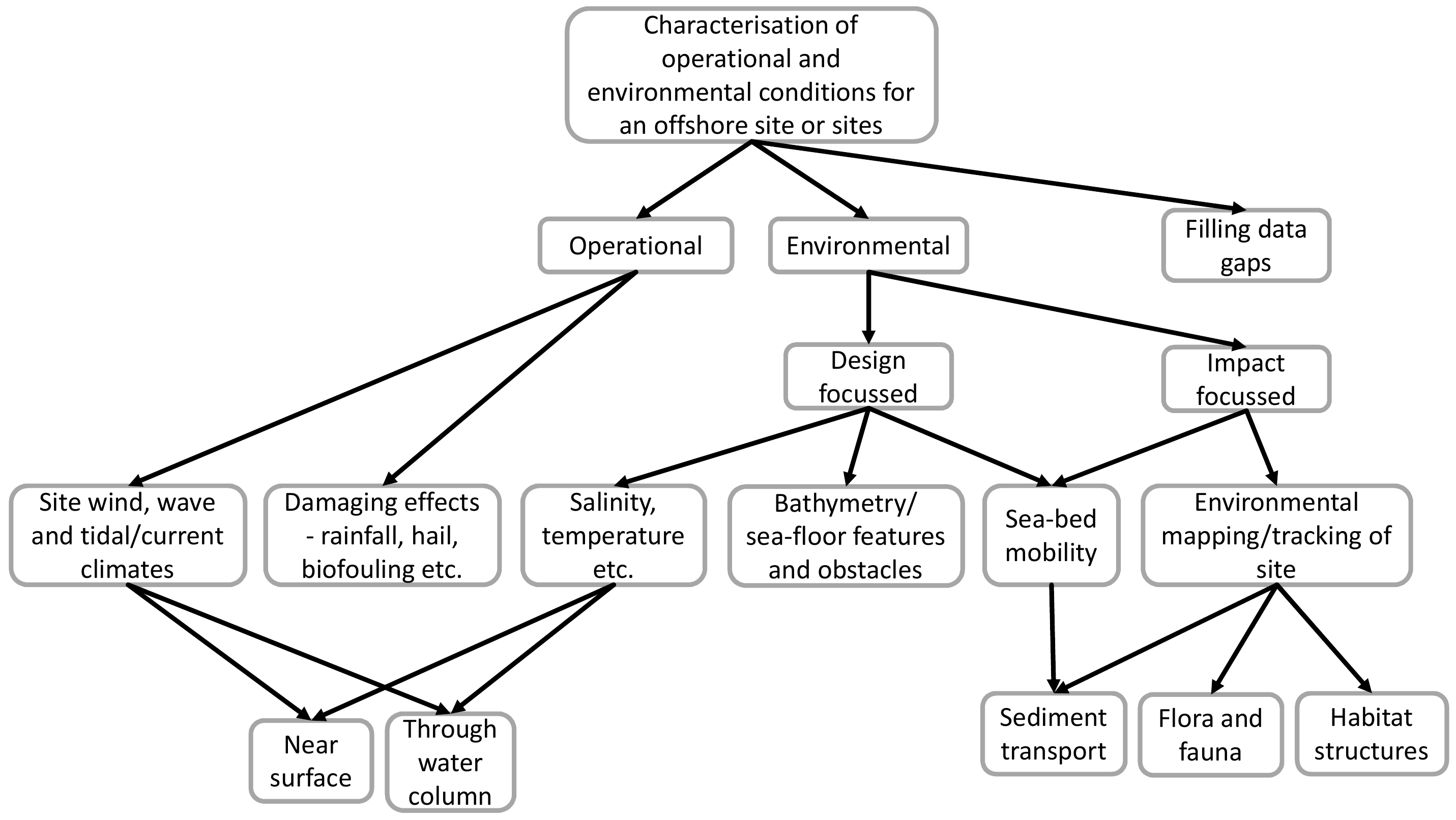}
\caption{Graphical summary of key knowledge/decision points for offshore renewables regarding the characterisation of operational and environmental conditions across an offshore site.}
\label{fig:summaryFig1}
\end{figure}

\begin{figure}[!h] 
\centering
\includegraphics[width=0.95\textwidth] {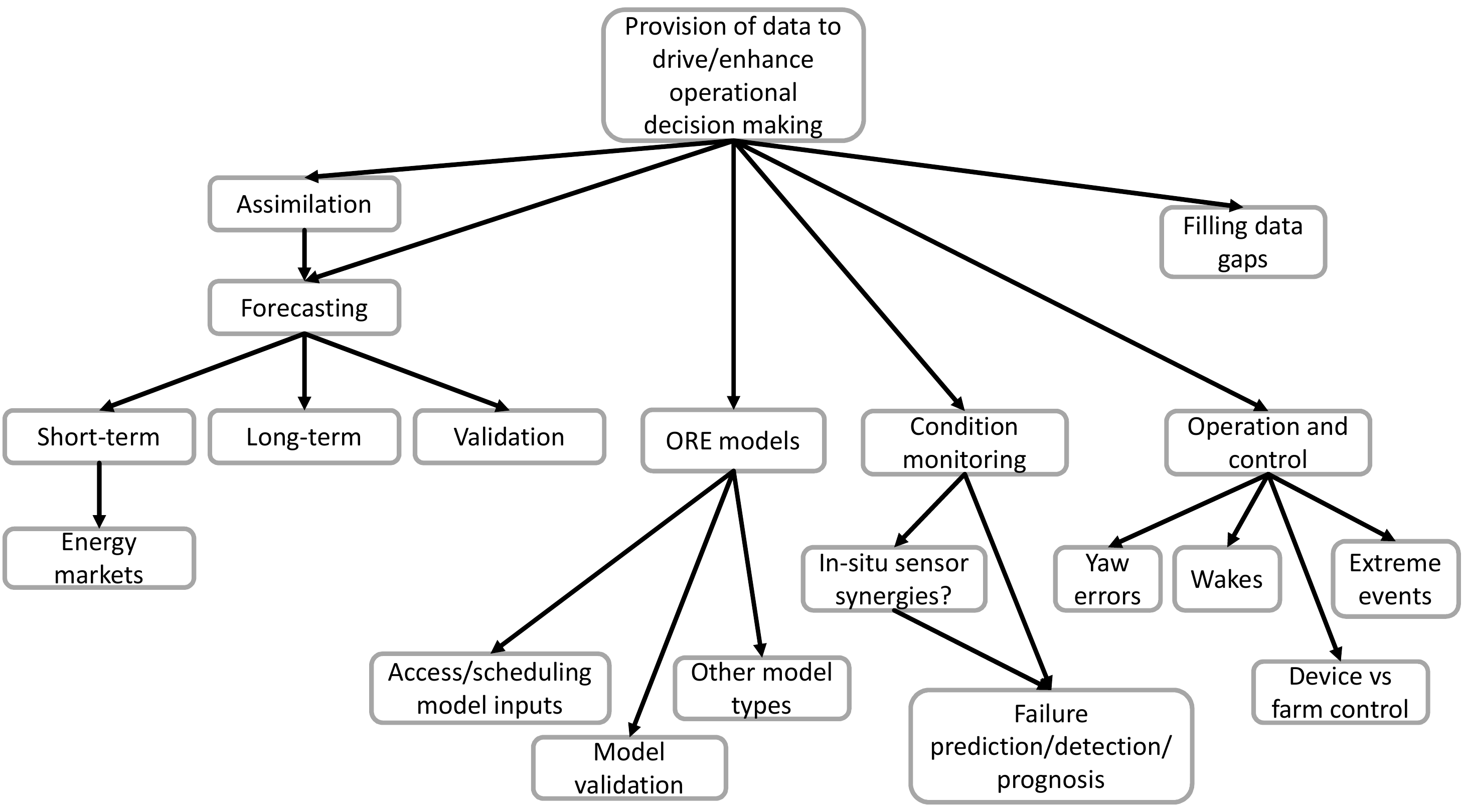}
\caption{Graphical summary of key knowledge/decision points for offshore renewables regarding the operational phase of an offshore site.}
\label{fig:summaryFig2}
\end{figure}

The current section introduces wind, tidal and wave energy, before discussing key innovation points and opportunities for which satellite data might add significant value. Within these discussions, it will become clear that commonalities exist across different ORE sectors regarding the decisions being made and the applications for which additional data is required. At a high level, the ORE sector has been found to broadly require improved information in two contexts: \emph{the characterisation of operational and environmental conditions for an offshore site} and \emph{provision of data to drive/enhance operational decision making}. Each category can be broken down further into a general framework within which key decisions across ORE technologies fit. These more detailed breakdowns are shown in Figures \ref{fig:summaryFig1} and \ref{fig:summaryFig2}.

\subsection{Offshore wind energy}

\subsubsection{Brief history of offshore wind energy}

The worlds first offshore wind farm was built in Vindeby in Denmark in 1991 \cite{ptarticle}. Commercial offshore wind began to gain some momentum in the early 2000s as a reaction to volatile fossil fuel prices and climate change, with the UK commissioning a 4MW test site at North Hoyle in 2000 \cite{owfGuide}. This was followed in 2002 by the first large scale ($>$100MW) offshore wind farm at Horns Rev in Denmark. Since then, offshore wind capacity has steadily increased worldwide to around 30GW today. Offshore wind energy is critical in meeting the climate aims set by 2050, although their realisation will require a more than ten-fold increase in offshore wind power \cite{IRENA19} compared to currently installed capacity. As such, significant innovations and costs reductions which are sustainable in the long term are required throughout the life-cycle of offshore wind farms \cite{BEIS19}. The following sections highlight the context for offshore wind decisions across the lifetime of a wind farm and indicate those areas where valuable contributions can still be made.

\subsubsection{Site selection}
The resource available at a site is a key parameter in determining the profitability of a resulting wind farm. In addition, site conditions will correlate strongly to asset lifetimes, for example high levels of turbulence are associated with early fatigue related failures. As such, more detailed information \emph{a priori} about a given site allows for more accurate yield and fatigue damage assessments to take place early in the project lifecycle. It is currently standard to assess wind conditions at a regional level, and once a site has been chosen at that fidelity an \emph{in-situ} campaign is undertaken at the site itself \cite{Semp08}. While the regional level analysis does often include satellite data of some form, a clear innovation which would add value to offshore wind site selection processes would be the possibility to survey much larger sections of ocean with increased detail and accuracy to better identify promising sites. Furthermore, site selection is now beginning to be considered within wider contexts which rely on availability of a range of data types \cite{Vagiona18,Cradden16} and GIS software analyses, and so again there is significant room for value to be added within these more comprehensive procedures and demanding data requirements.

\subsubsection{Wind farm design}
The design of a wind farm is a broad and complex task which ideally considers each stage of the plant lifecycle, including: technology selection \cite{Arram19}, layout and cabling optimisation \cite{Hou19} as well as operational and end-of-life practicalities. It should be highlighted that this is a key stage from an overall costs viewpoint, since, decisions taken at the design stage effectively lock-in certain cost outgoings for the full 25+ years lifetime of the wind farm. Decisions at this stage are very much driven by past, current and future site conditions and physical properties in terms of resource, additional environmental factors such as wave and current loading, sub-sea features and sediment type. Within the decisions to be made, substructure selection is a point where the project developer has strong influence since this choice is not controlled by the original equipment manufacturer (OEM). This decision in particular is heavily influenced by environmental and geological site conditions \cite{foundDesign15}. Uncertainties regarding site conditions currently lead to large safety factors being applied which inevitably add costs onto projects, better information which allows for these margins to be safely slimmed down would have important cost saving implications.

\subsubsection{Construction and installation} 
Offshore wind farm construction and installation requires expensive heavy lifting vessels to be utilised, with specialist equipment for foundation construction or piling also required. The use of such vessels is a significant costs to any project. They must be booked in advance and hence unused operational days are still charged at the normal rates. As such, careful planning and confidence in available weather access windows are crucial at this stage. In addition, vessels routes both to and within the wind farm, \emph{i.e.} the order of construction, must be optimised in order to avoid wasted fuel and additional costs \cite{Barlow15, LAR19}. Crew safety is also of paramount importance during these phases.

\subsubsection{Operations and maintenance}
The operations and maintenance phase of an offshore wind farm life-cycle is best understood as consisting of three distinct periods, these being: in-warranty operation, post-warranty operation and late-life operation. In-warranty operation last for the first 3-5 years once a site begins energy production. Within this phase the owner has limited control over how the site is run, depending on how contracts are arranged, and weather associated risks can in some cases be outsourced to the OEM. The agreements secured in these maintenance contracts have important implications for site yield \cite{Hawker15} in the early years. Post-warranty, site operators can choose to sign a new long term contract with the OEM, appoint a third party or bring these operations in-house. A mixture of the above can also be used. At this stage it is crucial for the site owner to understand the weather risks associated with operations, reliability and access. Key decisions centre around the planning and scheduling of maintenance and repair operations throughout the wind farm, while accounting for and managing the risks associated with weather and site conditions, as well as the stochastic nature of failures \cite{Seyr19,Shafiee15}. Understanding and monitoring Key Performance Indicators (KPIs) is also important at this stage \cite{Gonzalez17}. It could be argued that site conditions and weather risks for a given site should be well known after 5 years of data collection and analysis, however, with the increasing scale of offshore wind farms (hundreds of wind turbines covering hundreds of square kilometres of ocean) and often only a handful of on-site met-ocean sensors, detailed site data with a high resolution of spatial coverage is not yet a reality.\\

Late-life operations, including any periods of life extension, sees another shift in context for operational decision making. At this stage of the project, assets will be wearing out and failing with greater frequency and inspections and repairs will be required at shorter intervals. Furthermore, since Contracts for Difference (CfD) agreements generally have a life of 15 years, the site income may well have changed from those under which it was designed. CfDs are contracts that guarantee a payment for any differences between a ``floor price" (below which the market price of electricity cannot fall without a compensation made) and the actual cost of electricity to the provider. The end stage of the CfD agreement will therefore be driven strongly by the price floor. For a detailed overview of the key considerations and influencing factors for O\&M decisions and planning see \cite{Seyr19rev}.

\subsubsection{Decommissioning and life extension}
The decisions made towards the end of the life of an offshore wind farm overlap strongly with those made during the design phase, since, it is now necessary to reconcile assets' operational histories with what was assumed during planning and design. Decommissioning costs themselves are also somewhat uncertain and so reductions in uncertainty around this is important for optimal decision making \cite{Topham16,Topham19}. Life extension of a wind farm consists of assessing whether the assets can be safely and profitably operated beyond the end of their design lives, a question which would be greatly assisted by improved site condition information across the wind farm lifetime. These decisions have been fairly extensively studied in the context of onshore wind \cite{Zieglerrev18}, with offshore now also being considered \cite{Bouty17}, albeit at an earlier stage of understanding and experience. 

\subsubsection{Future challenges for the offshore wind sector}
Innovation points and the possible provision of information which would have significant impacts to the offshore wind industry all centre around the provision of more data, and with improved spatial resolutions in particular, over what is currently available regarding site met-ocean and environmental conditions. In addition, the enhancement and increased redundancy of offshore wind site communications provisions, for the purposes of providing ancillary services and ensuring security of supply/blackout prevention (with respect to a future grid with much higher proportions of renewables penetration), would also have a significant impact in this sector.  

\subsection{Tidal energy}

\subsubsection{Brief history of tidal energy}
The use of tidal energy  dates back to as early as 10AD (first recorded tidal mill on the Persian Gulf)\cite{Charlier2009}. Early engineering  applications exploited both tidal stream and tidal range energy to turn water wheels for milling grain. Investigation into the generation of electricity at a large scale  began in the 1920s, with the world’s first tidal barrage power station operating at La Rance \cite{Neill2018,Khare2019}, France, from 1966. The La Rance barrage remains operational, generating power through an installed capacity of 240MW. Interest on the generation of electricity from tidal stream began in the 1980’s when research into the design of tidal hydrokinetic energy converters \cite{Bahaj2003,Laws2016,Borthwick2016} started to become mainstream. A variety of different rotor designs have been trialled, but the sector has converged toward the horizontal axis tidal turbine (HATT), \cite{Bahaj2003}, for the majority of applications, following the design technology developed for the wind sector. To date power generation from tidal stream has not reached the level of tidal range systems in energy output terms. The largest power producer from tidal stream is the MeyGen Tidal Power Project situated in the Inner Sound of Pentland Firth, Scotland. MeyGen is currently extracting 4.5MW through an array of four turbines, with resource consent to extend this to 86MW, \cite{atlantis}. 

\subsubsection{Tidal resource assessment and site characterisation}
Tidal energy is generated by the gravitational interaction of the Earth--Moon--Sun system \cite{Gill1982,Pugh1987,Apel1988}. As the relative orbital positions of these three bodies evolves, the gravitational and centrifugal potential at the Earth's surface varies. This leads to deformation in fluids on the Earth's surface; the fluid effectively bulges at nodal points in response to local changes in gravitational potential - this is the tide \cite{Pugh1987,Apel1988}. 
Tidal dynamics vary in time due to the interaction of multiple forces. These tidal forces can be described by mathematical expansions of harmonic constituents, a more detailed description can be found in \citet{parker2007tidal}. The period difference in these forces correspond to longer spring-neap patterns, as well as other cycles that contribute to tidal range and velocity variability. There are certain periods when the tidal forces constructively interfere to increase the tidal range. This applies over spring tides when the principal lunar ($M_2$) and solar ($S_2$) forcings complement each other resulting in greater than average tidal ranges and velocities. The opposite occurs during neap tides when forces from the moon and the sun counteract one-another. Moreover, other hydrodynamic effects, as ell as wind and waves, \cite{Lewis2017}, will also be relevant to the evolution of the tide,  altering the available potential and kinetic energy.  \\

Apart from temporal variations, tides also vary spatially. Tides are by definition ocean surface waves of a very long wavelength $\lambda$. Despite the pronounced wavelength, they remain subject to wave transformation including shoaling and reflection.As a progressive wave propagates from deep water to shallower regions (e.g. in the transition from oceanic depths to a continental shelf), the wave celerity $c$ ($\approx \sqrt{gd}$, where $d$ is the depth) decelerates, reducing $\lambda$ and leading to an amplification of the wave height $H$. In combining such changes in depth with further reflections whilst encountering coastal features,  reflected and incident tidal waves combine to form a perceived total wave \cite{pugh2014sea}. Consequently, the presence of landmass on the Earth's surface higher than the mean sea level introduces obstacles to the propagation of the fundamental tide (i.e. the tide main component), resulting in local distortions in the tidal amplitude, phase, and harmonic constituents \cite{Pugh1987}. It is these local distortions (for example, interactions between islands, or narrow sea channels between land masses) that lead to the energy amplifications exploited for tidal energy extraction.\\

The site information required by a tidal developer depends on the method of energy extraction being developed, i.e. tidal range or tidal stream. Tidal range systems take advantage of local resonance effects that amplify the available potential energy over a tidal cycle. Resonance in a tidal basin occurs when the tidal wave period $T$ is close to the natural period of the basin, as for example given by Merian's formula \cite{Pugh1987}. Effectively, the amplification of an incoming wave is increased if the basin length converges to $\lambda/4$, assuming idealised bathymetric conditions. Such resonance effects occur in several sites across the globe. 
In particular, the highest tidal range has been observed in the Bay of Fundy, Canada where the basin natural frequency has been calculated to be $T_b \approx 13.3$h leading to amplifications \cite{garrett1972tidal} for the $M_2$ and $S_2$ tidal wave constituents. In the UK, equivalent conditions emerge within the Severn Estuary, known for hosting one of the highest tidal ranges in the world. As a result both of the above have been the focus of many tidal range energy proposals.\\  

Tidal stream systems take advantage of localised flow acceleration resulting from flow constriction (channels and straits) or flow curvature (headland effects). Such features impede locally the propagation of the tidal waves, resulting in hydraulic gradients. The latter amplify current velocities, and by extension the scale of the kinetic energy available.  
For all forms of tidal power generation, resource assessment is typically a combination of measurement and numerical modelling; this is the approach recommended in the IEC technical specification, \cite{IEC_TS62600201_2015}, for tidal energy resource assessment and characterisation.\\

Tidal range energy resource assessment initially relies on the potential energy calculation. The maximum theoretical potential energy produced by the head difference $H$ (m) is given by \citet{Prandle1984}: $E_{\textnormal{max}} = \frac{1}{2} \rho g A H^2$. $\rho$ is density, $g$ is acceleration due to gravity, and $A$ is the impounded surface area.  Using this basis, the estimation of the global resource has been the subject of recent studies (e.g \cite{Neill_review2018}) that employ tidal elevations drawn from available global tidal models to quantify the head differences over every tidal cycle. While global ocean models can be a useful indicator to provide a coherent insight to the distribution of potential energy,
these models typically lack resolution in the coastal zone and exclude critical processes such as intertidal effects.
Therefore, global approaches may not capture localised coastal details that can be vital to reproduce the resonance effects accurately. In refining the resource estimates for individual assessments, regional coastal ocean models are sequentially applied to quantify with more confidence the regional conditions \cite{Angeloudis2017}. In conducting these processes reliably, measurements become essential to accurately capture the tidal elevations and velocities in the areas of interest, ensuring that the numerical approaches adequately reproduce key dynamics. In particular, for a more informed resource and impact assessment of more developed schemes, operational models are coupled with coastal ocean models \cite{Angeloudis2016} to simulate the individual scheme operation and quantify and optimise the expected energy output \cite{Angeloudis2018}.\\

The most basic tidal stream resource assessments aims to determine how much kinetic energy is available for extraction at a given site. The instantaneous power at a given point in the fluid is defined \cite{Bryden2007} as $P_{inst} = \frac{1}{2} \rho U^3$, where $\rho$ is the fluid density, and $U$ is the horizontal fluid velocity. To calculate the available tidal stream resource a 3-D map of the time varying fluid field is required. Energy is extracted from a vertical plane in the flow corresponding to the rotor plane of a tidal turbine. The available power is the area integral over the rotor plane of the power from the horizontal flow normal to the rotor plane, integrated over time. The converted energy extracted depends on the power efficiency ($C_{P}$) of a given turbine. This analysis does not take into account the impacts of levels of turbulence, locally generated large-scale flow structures (\textit{e.g.} eddies), or local wave-current interactions. All of these processes impact on a TEC's ability to extract energy from the tidal stream and its long-term reliability. Therefore resource assessment for tidal stream energy requires a full site characterisation. This informs where best to locate turbines and turbine design parameters, as well as higher frequency loadings.\\

\begin{figure}[!ht] 
\centering
\includegraphics[width=0.85\textwidth] {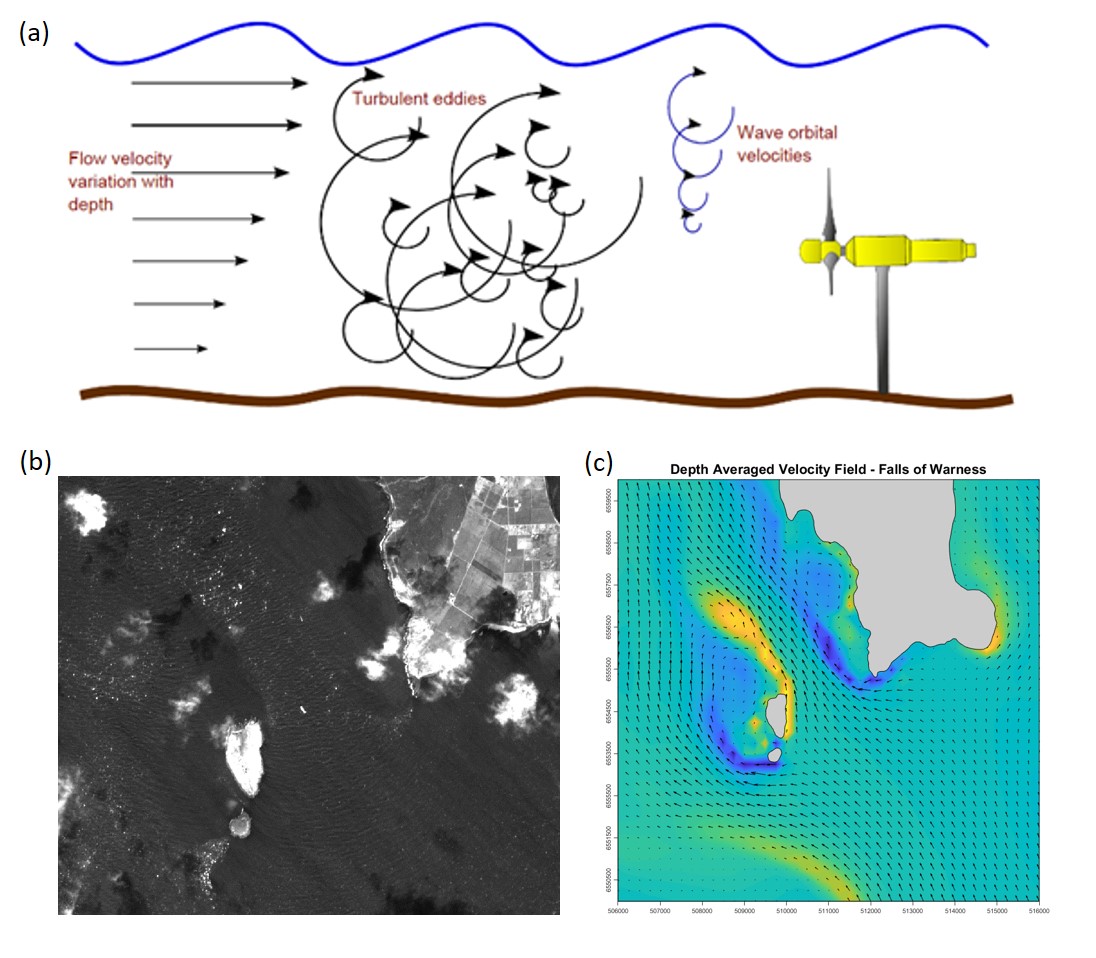}
\caption{Site characterisation for tidal stream energy extractions needs to capture the full range of complex flow conditions. (a) The IEC/TS62600201\_2015 \textit{in situ} data specifications target the capture of the vertical variation in flow conditions, (b) but the spatial variations due to flow separation processes, as observed in satellite imagery (Sentinel2A channel B3 data, 21/04/2017), need to be considered, (c) this is where regional-scale modelling is required to extend the information content. The lower pannels show the EMEC research site.}
\label{fig:tidal_stream}
\end{figure}

Site characterisation is achieved through a combination of \textit{in situ} measurements and regional fluid modelling. \textit{In situ} measurements are provided from a variety of instruments, the most common being the Acoustic Doppler Profiler (ADP). An ADP consists of multiple acoustic transducers pointing in different off-axis directions, but in a geometric pattern that allows the reconstruction of the 3-D velocities along the instrument axis. They measure the fluid velocity through the back-scattered Doppler shift in the frequency of a transmitted acoustic pulse. Details of these instruments and their operation can be found in \cite{Thomson2014,Joseph2013}. ADPs can be deployed in fixed seabed moorings, mid-water or surface floating moorings, or from a moving vessel. In high-flow regions of interest to tidal developers, only the seabed mounted and vessel mounted measurements are made, due to the harsh environmental conditions. Seabed moored instruments provide a time series of vertical profiles of the 3-D velocity at a single point. These give high temporal data, but only the vertical flow structure at the mooring location. Vessel mounted ADP's can be used to survey a series of transects over multiple tidal cycles, providing good spatial coverage at the expense of temporal resolution. To capture the full spatio-temporal variability in the flow field, 3-D regional tidal models are required. An appropriately chosen set of ADP measurements are used to validate a regional model. When appropriately configured, the ADP's also provide measures of the local turbulence, \textit{e.g.} turbulence intensity ($TI$), turbulent kinetic energy ($TKE$), turbulence dissipation rate ($\epsilon$), or components of the Reynold's stress tensor, and they can be used to determine turbulence length scales. All of these parameters are relevant to turbine design and reliability modelling. Lastly ADP measurements can be used to quantify the local wave-current interactions. Wave measurements are made using the methods to be described in the section on wave energy. At the time of writing the IEC provide technical specifications \cite{IEC_TS62600201_2015} defining data collection and processing, numerical model configuration requirements, and site characterisation reporting methods for site developers. \\

Advanced measurement methodologies are being developed to collect site data from operating turbines \cite{Draycott2019}. These techniques provide opportunities to directly assess the impact of a turbine on the local flow and to complement standard measurements that have to be made up-stream and down-stream of a turbine to minimise instrument deployment and recovery risks. These data allow the quantification and assessment of theoretical parameters used to define the IEC technical specifications for data capture. Local velocity field ``gusts" and their impact on turbine behaviour can be directly measured through the use of horizontally mounted ADPs. Bespoke acoustic instruments are being designed and tested that will provide more accurate measures of the hub-height flow structures, than the standard off-the-shelf instruments can currently achieve. The integration of such instruments on turbines may also provide a means for potential responsive control of turbines to optimise their operation and minimise fatigue. \\

\subsubsection{Main tidal technologies}

Tidal range power plants behave effectively like dams, constructed in areas exhibiting sufficient tidal range to economically house turbines for power generation. Their operation is based on the creation of an artificial tidal phase difference, by enclosing water over a surface area $A$. 
This facilitates a head difference $H$ that is then allowed to drive flow through turbines, tapping into the potential energy. Tidal range power generation is a low-head hydropower application where bulb turbines are currently the default turbine technology proposed and installed.
Over the years, bulb turbine capabilities have been continually evolved in terms of their efficiency \cite{Waters2016},  but also in terms of their sustainability. In terms of the overall tidal power plants, while the constituent elements between designs remain the same (i.e. \textit{Turbines, Sluice gates, Embankments}), the configuration may vary to deliver considerably different generation profiles. \\

\begin{figure}[!ht] 
\centering
\includegraphics[width=0.8\textwidth] {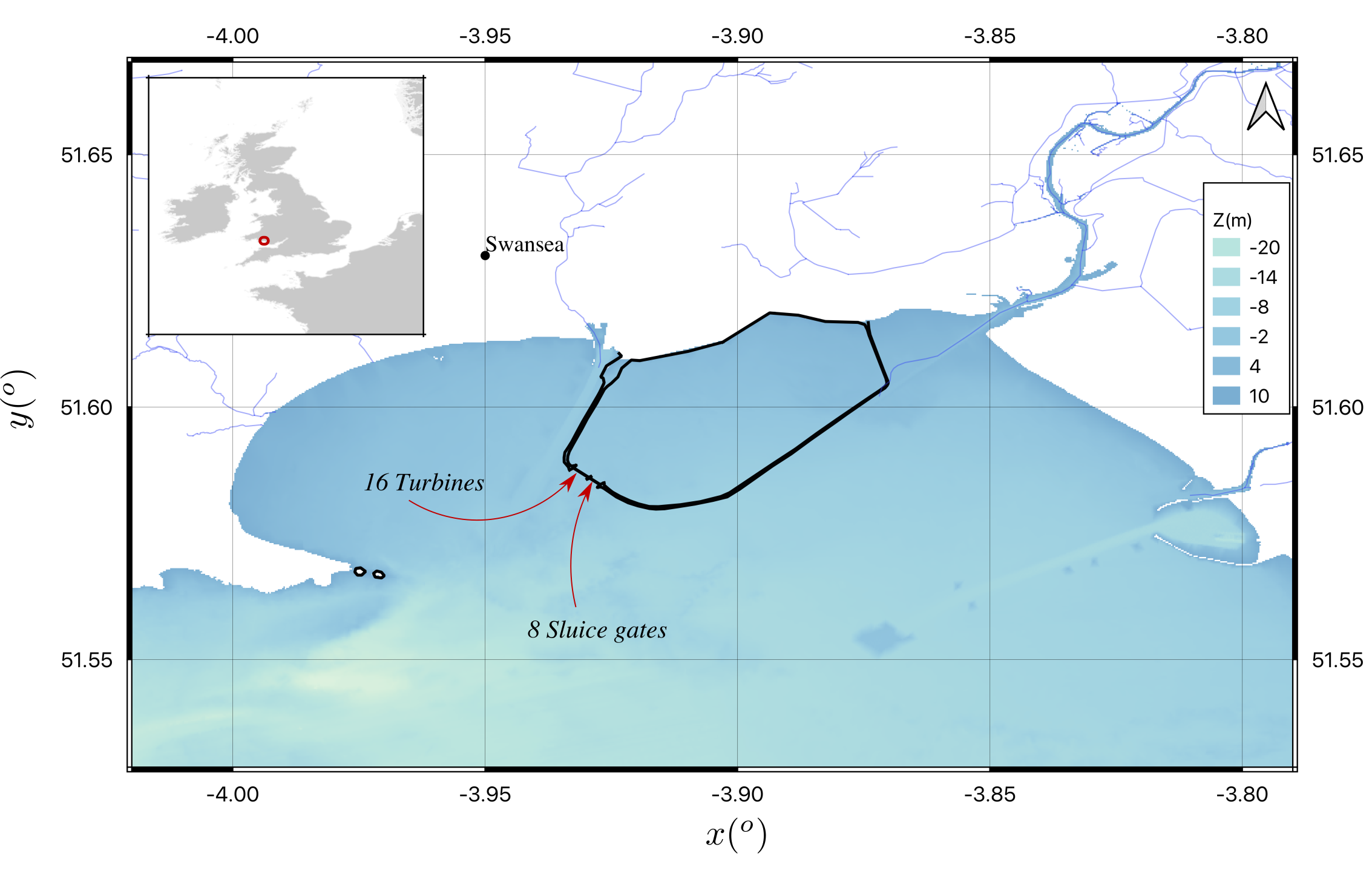}
\includegraphics[width=0.8\textwidth] {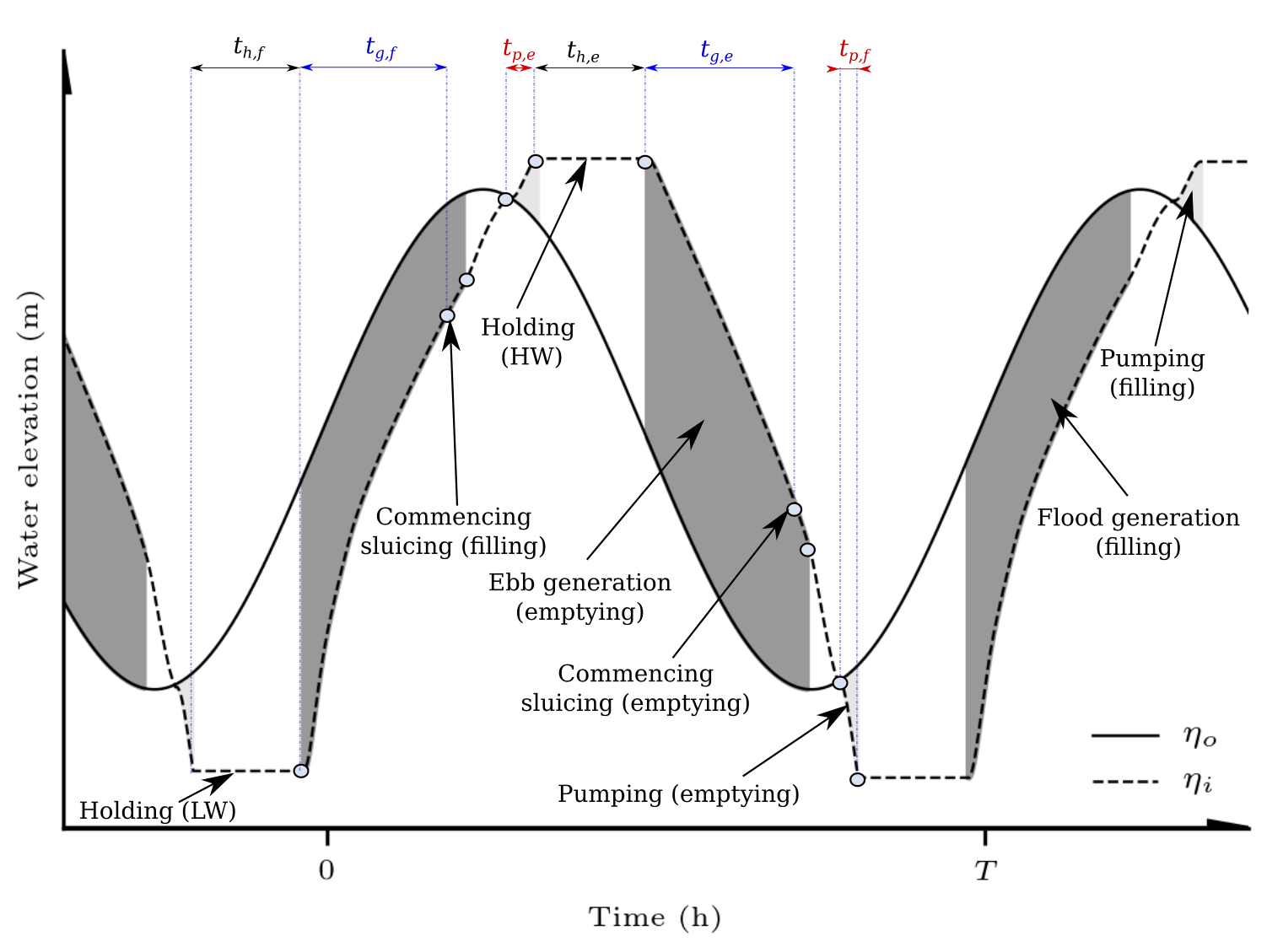}
\caption{General arrangement and operation of tidal range power plants. Top: the Swansea Bay Tidal Lagoon outline of Tidal Lagoon Power Ltd, an example of a single-basin tidal range structure. Bottom: Operation of a tidal range power plant including control parameters that can be used to optimise a tidal range structure operation \cite{Harcourt2019}.}
\label{fig:tidal_range}
\end{figure}

Tidal range power plants can be attached to the coast (such as a barrage or a coastal lagoon) or located entirely offshore (as offshore lagoons). 
`Barrage' and `lagoon' are terms used as per the impoundment perimeter characteristics. 
If most of the perimeter is artificial, then the term lagoon is appropriate, whilst barrage is suitable otherwise. 
In the case of lagoons, the additional cost for longer embankments typically enables smaller developments, that may be more feasible than barrages. The latter  typically span the whole width of an estuary, and are thus associated with severe hydro-environmental impacts that are challenging to quantify.\\

Tidal stream energy extraction is still an emerging  technology, that has accelerated in recent years due to the need to reduce carbon emissions to mitigate the impacts of energy generation on climate change. The principles of tidal stream energy extraction are very similar to those for wind: a rotor system is coupled to an energy converter that generates electricity which is supplied to a power grid. The majority of marine tidal turbines being developed are horizontal axis systems. There are two mooring systems currently being investigated: fixed bed-mount systems such as those from MeyGen and Nova Innovation, \cite{atlantis, nova}, or floating systems, such as the Orbital Marine turbine, \cite{orbital}. Research to date indicates that the fluid dynamics in tidal stream energy extraction sites are significantly more complex than those encountered by the wind sector, therefore some of the assumptions that have been carried across from wind may not be valid. This has tended to lead to over-engineering of tidal turbines with an associated increase in production costs. There is a large amount of on-going research investigating how to provide more appropriate information to both designers and site developers to help improve turbine performance, reliability, and maintenance. Beyond the single turbine systems, there is ongoing research into the design of tidal turbine arrays, in particular turbine interaction and optimisation of power production through choice of array layout. There is also an associated impact on both the local environment and away from the array due to the extraction of tidal energy. These impacts are still not fully understood, and are a subject where there is further potential to take advantage of satellite data to capture spatial changes over a wide area.\\

The development of tidal energy sites is both a difficult and costly exercise, as these site inherently have very strong and highly turbulent flows. Marine operations are limited to narrow windows around times of slack water (\textit{i.e.} there are four operating windows each day at times of high and low water). The operating windows are further reduced by weather and waves (both locally wind generated and open-ocean storm generated swell). For these reasons, tidal energy extraction systems are costly to install, maintain, and operate. To be economic they must be scalable. To achieve the necessary level of scalability high quality site information is required, including extreme event statistics. Both short-range and long-range forecasting is required to help minimise operational and maintenance costs.

\subsubsection{Tidal projects}
The majority of large scale operational projects are of the tidal range type. Up to July 2020,
 the largest tidal energy producer is the Sihwa Lake Tidal Power Station in South Korea, with a 254MW capacity \cite{Bae2010}. This is followed by the La Rance Tidal Power Plant, France (240MW) \cite{OESReport2019,Waters2016}. Smaller tidal range schemes exist in Canada (e.g. the 20 MW Annapolis Royal tidal power station), Russia and China. Moreover,  there is interest in developing schemes at multiple sites where the resource allows it \cite{Neill_review2018}, or as  solutions supporting electrification in remote/isolated areas \cite{Mejia-Olivares2020,DELGADOTORRES2020114247}. 
Scotland is at the forefront of the developments in terms of tidal stream projects. SIMEC Atlantis and the Meygen array are devising 4.5 MW of tidal stream energy in the Pentland Firth, \cite{atlantis}. Nova Innovation is developing a tidal array in Shetland, with a capacity of 300 kW, \cite{nova}. Orbital Marine Power has developed and tested a 2MW turbine at the European Marine Energy Centre (EMEC) in Orkney, \cite{orbital}.

\subsubsection{Challenges facing tidal sector}
Tidal range energy technology is generally perceived as a mature form of power generation. This is attributed to successful tidal power plant examples operating around the world, as well as the technology transfer from hydropower and wind energy sectors, both long--established technologies. Nonetheless, there are several factors  dictating the development of tidal energy schemes even if preliminary resource assessments indicate commercial potential \cite{Neill_review2018}.
Beyond physical constraints, the capital cost and uncertainty over the environmental impacts of large marine infrastructure constitute formidable barriers to development.\\

The presence of an obstruction such as a tidal range scheme can lead to earlier reflection of the tidal wave and thus interfere with the established tidal resonance. 
The impact of the structure is more pronounced for barrages that typically span the entire basin. 
As a consequence, the tidal range at a regional scale can be altered, with implications for tidal fluxes, coastal inundation and the size of intertidal areas. In the case of tidal power plants positioned along a coastline, advective accelerations can take place as the flow goes around the confined area \cite{Angeloudis2020}. 
Changes in velocity magnitude modify the shear stress exerted on the seabed, thus leading to recirculation zones \cite{Vouriot2019} triggering erosion/deposition processes, with potentially long-term results.\\

Impacts on sedimentary processes can sequentially alter estuarine and coastal water quality. 
Increased sediment deposition can lead to a reduction in Faecal Indicator Organism (FIO) concentrations (e.g \textit{E. Coli}), thus fostering improved water quality, subject to the interaction of the designs with the hydrodynamics and other existing marine infrastructure (e.g. sewage or desalination inflows or outfalls).  
The increased residence time and light penetration in the artificial impoundment could favour conditions for primary phytoplankton production.  Salinity concentrations are likely to change, particularly in the impounded area in the presence of substantial brackish/fresh water sources.
A more extensive discussion on the water quality impacts of tidal range developments can be found in \citet{Kadiri2012}. \\

The main challenges facing tidal stream development are related to turbine reliability, reduction of cost through optimal design, optimal planning of operations and maintenance activities. Reliability can be improved through an improved understanding of the fluid dynamics, in particular quantification of the large-scale structures, wave-current interactions and their effect on turbine behaviour, and predictive control of the turbine to minimise impacts of high-frequency events. The design of turbine rotors has a relevant impact on the operation and reliability of tidal turbines. Blade design is informed through Computational Fluid Dynamics (CFD) modelling. The CFD model needs to provide true representations of the full fluid environment (i.e. currents, small- and large-scale turbulence, and waves) to capture the full spectrum of the physical loads. Operations and maintenance is costly in the marine environment, particularly for sea bed-mounted turbines.  \\

To date, site characterisation has been predominantly carried out using \textit{in situ} measurements and numerical models, but satellite data provide regional scale data the can be used to both characterise a site and validate numerical models. In particular, optical and SAR imagery can be used to quantify the large-scale eddy fields generated by flow separation processes. These data potentially provide another independent measure that can be used to validate numerical simulations, and provide a first pass indication of regions within a site that may be problematic both for \textit{in situ} data collection for site characterisation and for subsequent positioning of tidal turbines. Optical imagery generally provides high spatial resolution (<10m), but the information content is limited by the amount of cloud in a given image. This significantly reduces the number of useful images available for a given location from a time series. SAR imagery is not restricted by the presence of cloud and the spatial resolution is improving with each generation of instrument. With an appropriately configured SAR system it has been shown that surface currents can be measured from space \cite{Joseph2013,Romeiser2014,suchandt2015}. These techniques need development to allow full 2D surface-velocity fields to be derived, but has the potential to provide maps of surface flow data. Altimeter data are used to validate tidal models, but currently these data do not provide sufficient spatial resolution or accuracy in straits and channels typically targeted for tidal stream energy extraction. The more general limitation of satellite data is the frequency of over-passes at any given site, which is generally twice daily or less, but there is the potential to build a statistical picture of temporal variability from long-time series of satellite data.

\subsubsection{Future of tidal energy}
For tidal range energy, there are a number of schemes that are under consideration over the next 5-10 years, and many more sites have been  identified around the globe where  schemes can be developed. The main barrier to the development of these types of systems is generally related to environmental impacts as well as making a robust case for the initial capital investment. For the latter, satellite data can play a role in more adequately addressing uncertainties associated with the environmental impact and also informing future studies about regional and far-field effects of tidal range schemes.For tidal stream, increased commercial investment is required, most of the current focus is on small scale projects that service islands or integrate with a range of renewable energy sources. Reliability is a key factor identified as introducing high-risk to investors, so significant development is needed to accelerate methods for improving reliability. This depends on design improvement, materials, turbine siting, resource characterisation, predictive control (impact of waves, storms, etc), and O\&M optimisation.

\subsection{Wave energy}
 
 \subsubsection{A brief introduction to wave energy}
 
Over many years, significant research, development and innovation has been carried out across the world looking at wave energy technologies. The first recorded attempt at devising a concept to harness wave energy goes back to 1799 when two French engineers patented a system involving a float and a lever \citep{RossDavid1995Pftw}. There is however, no knowledge of the outcome of this invention. The next notable use of wave energy was in the 1950s to power navigation buoys around the coast of Japan. These buoys were produced on a relatively large scale (about 300 were deployed) but the amount of energy each buoy produced was relatively small (just enough to power a 60W navigation light). Modern research and development in wave energy as a means of producing utility scale electricity was motivated by the oil crisis of the early 1970s. New government programmes, particularly in the UK, encouraged research, including  pioneering work of Prof. Stephen Salter and his team at the University of Edinburgh \citep{salter-74}. Since then, over 85 wave energy converter (WEC) concepts have been tested at large-scale and under representative environmental conditions \cite{Babarit2017}, but so far, no long lasting commercial breakthrough has been achieved. Through changing programmes at different levels of government, support for technology development was intermittent and progress slow \cite{Hannon2017}. At the same time, other forms of renewable energy have been progressing quickly, taking some focus off the developments required to push wave energy technology forward. As a results, some technology developers are now focusing on niche markets such as powering remote island communities, oceanographic instruments and isolated offshore oil and gas equipment, where traditional power generation approaches are costly and/or difficult to implement technically \cite{LiVecchi2019}.\\

Over the course of wave energy development, a wide variety of methods have been explored in an effort to extract the energy from waves. Each of these methods has tried to exploit different features of the wave energy resource. Some methods rely on the rise and fall of a wave moving a floating device, others rely on the change in pressure above a submerged device, while another method extracts energy from the surging motion of waves before they break close to shore. The Aquaret project \cite{aquaret} gives descriptions and graphics describing the most common types of WECs, each one designed around different features of the wave resource. Details relating to the performance of device types in different sites using a variety of evaluation metrics can be found in \cite{Babarit2012}.\\
 
A key performance indicator to assess the future potential of a WEC technology is the long term LCOE. Three key elements of this calculation are: the initial capital cost of the wave energy device/array, the operations and maintenance costs, and the revenue generated from output. Each of these elements are influenced at different stages of the project development lifecycle. The capital expenses can be considered to be loosely related to the historical site conditions, and the structural design of the WEC. Operations and maintenance (O\&M) costs can be minimised through design and planning but will be influenced by the site conditions at the time the O\&M operations are required. Indeed downtime for maintenance can be planned during periods when lost revenue would be minimised either based on analysis of historical data or forecasts of future conditions. Prediction of future revenue will be based on calculations of historical wave resource information and knowledge of device performance, but can be broken-down to seasonal revenue based on resource variations. Accurate assessment of the LCOE throughout the development stages of a WEC concept is key to assess its commercial viability and inform design decisions.

\subsubsection{Wave energy converter site characterisation} 
\label{}
The features of a site which are of interest to the developer of a WEC include, but are not limited to, seabed bathymetry and ground conditions, tides and currents, wave resource (wave heights, wavelengths, range of wave directions and spectral information), marine life and wind conditions. Other characteristics of the site which may be of interest depending on the type of WEC are water depth changes based on tidal range or variation between regular conditions and storm conditions. There are a range of instruments and methods available to determine some of these features which include \textit{in situ} measurement, numerical models and human observations. Each of these has their own advantages and disadvantages which include cost and accuracy, so the best compromise is often used for practical reasons. When technology starts to reach a higher level of maturity it is required to test a prototype at sea, and this is often undertaken at a scale which is smaller than the expected commercial device. The process to determine an appropriate open water site for a scaled device is not straightforward and detailed analysis of both the commercial deployment site and the scale test site is required \cite{Mclean2019}.\\ 

Improving the accuracy of available data for the site, including the wave resource, is imperative to optimise the design of the technology and the commercial array in which it is to be used. Regardless of the feature of the wave resource being exploited by the WEC, accurate representation of that resource is required. Seasonal and annual variations at a site can be significant \cite{Perignon2017,Mclean2019} so extensive knowledge of a deployment site is important at a number of stages throughout the development of a WEC and a commercial project. During design phases it is important to understand the wave conditions and how this will impact on the device motions and loads within individual components. The International Electrotechnical Commission (IEC), recommends, in its technical specification on wave energy resource and characterization \cite{IEC_TC114-101_2015}, fully spectral and directional description of the seastate for WEC design, as well as understanding of the extreme conditions at site in order to determine design load cases \cite{IEC2019}. This detailed level of resource description can be achieved by \textit{in situ} physical measurements or by hindcast numerical modelling. \textit{In situ} measurements can be obtained from a range of instruments. The most common ones are listed below with a short explanation and a more comprehensive list, with more details on their working principle can be found in \cite[chapters 3-4]{Tucker2001a} and \cite[chapters 3]{Pecher2017}.\\
 
 Surface-following buoys are equipped with accelerometers and/or GPS sensors from which wave elevation time series and spectral measurements can be inferred. These instruments are typically expensive (although new GPS buoys are becoming more affordable \cite{Raghukumar2019}). They are well-suited for deep water sites and have a well-established track record (less so for the GPS based ones which are more recent). Their main limitation is measurements where currents are present and steep waves as the buoys does not accurately follow the water surface. Directional and non-directional surface-following buoys are available. Regular removal of marine growth may be required to ensure measurement accuracy is not impacted by the change in weight or drag of the buoy. Acoustic current profilers are instruments that can be deployed on the seabed and which rely on transducers emitting acoustic pulses which are reflected by the water surface and by particles in suspension in water. The former phenomenon allows surface tracking while the latter allows measurements of particle velocity throughout the water column using the Doppler effect \cite{Work2008}. As their name implies, these instruments were originally designed to measure current but omnidirectional wave elevation can be measured from surface tracking while directional wave measurements can be inferred from water particle motions. These instruments tend to be expensive and suited for shallow water wave measurements. Pressure transducers are also deployed on the seabed. The pressure measurements they provide is used to infer the water level variations due to tide but also to waves. They tend to be less expensive but the quality of the wave measurements obtained is affected by the wave induced pressure decay with water depth which make them only suitable for shallow water sites. They only provide omnidirectional wave measurements except if they are deployed in arrays \cite{Howell1998}.\\
 
\begin{figure}[h!]
\centering
	\includegraphics[width=0.7\linewidth]{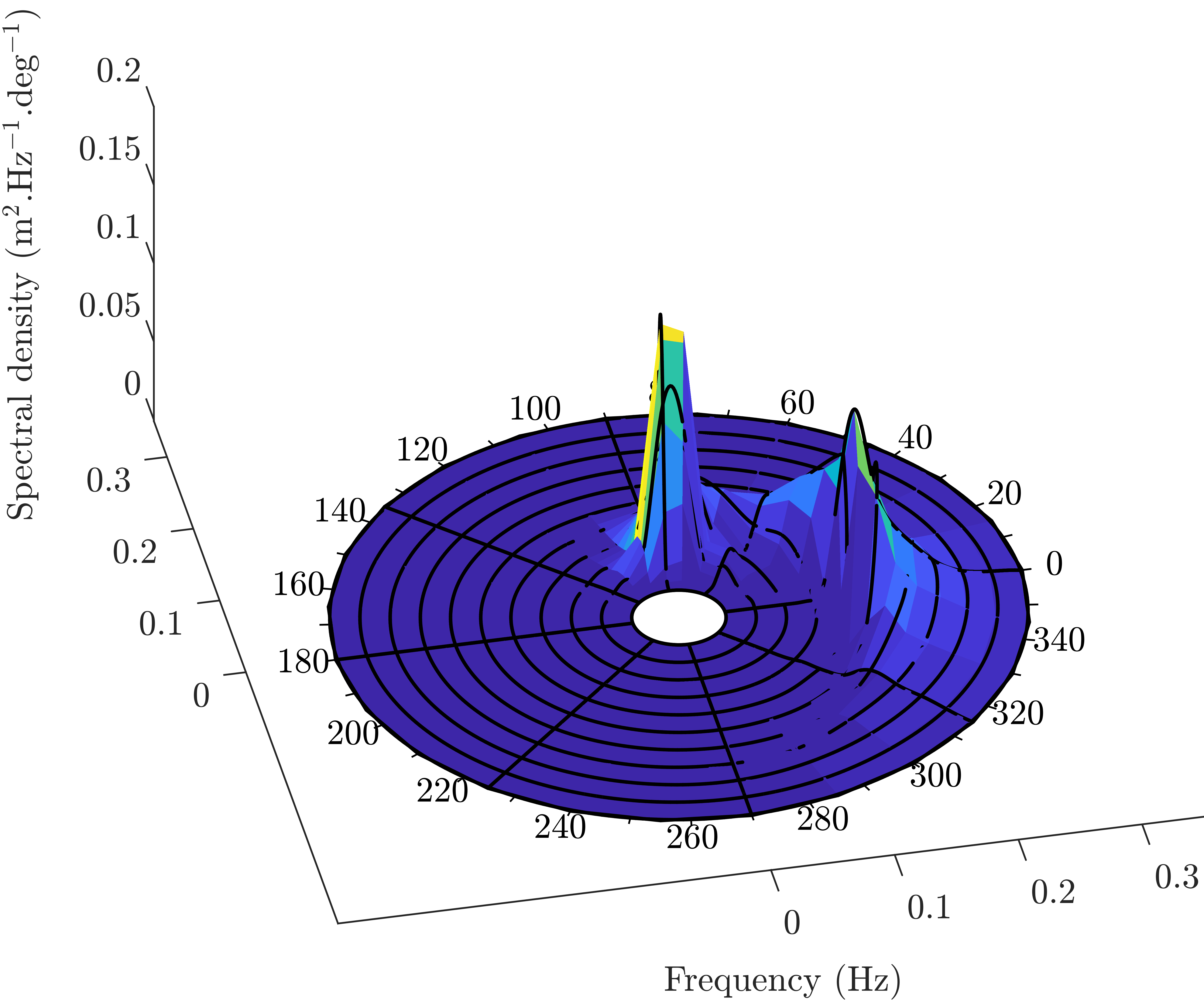}\\
	\caption{Frequency and directional spectrum at 47.239$^\circ$ North; 2.786$^\circ$ West (off the coast near Nantes, France).}
	\label{fig:spectrum}
\end{figure}

Numerical modelling of wave energy resource is typically achieved using third-generation spectral models such as SWAN or WaveWatch. These solve the wave action balance equation \cite{Hasselmann1988} on a grid which discretise the ocean domain of interest. More information about this is provided in section \ref{sec:wavesat}. Physically, this balance equation can loosely be assimilated to the energy balance of waves, where the wind blowing over the ocean surface is the main energy source and white capping, wave breaking and bottom friction are the main energy sinks. Moreover, the model takes into account non-linear energy transfer between different wave frequencies. The way some of these phenomena are modelled is semi-empirical and the model therefore requires calibration against \textit{in situ} measurements. It is not always possible to have location overlap between \textit{in situ} measurements, accurate spectral model data and desired WEC deployment sites, however it may be possible to use one buoy to calibrate a wave model at one precise location, then expand the area of acceptable accuracy based on satellite data. Figure \ref{fig:spectrum} shows a representation of the frequency and directional spectrum obtained from the public domain wave climate hindcast Homere database \citep{Accensi2015} produced by IFREMER.

\subsubsection{Wave climate forecasting} 
To optimise a WEC and an array, real time knowledge of the incoming wave can be used as an input to a control system to update system settings, \cite{elva1}. The optimisation may be designed for maximising power output, smoothing output across the array or intended to reduce wear on components. \\

Short term forecasting can also be used to estimate the output of the array, providing more accurate information to grid operators and analysts looking to balance the national grid. Balancing the grid is an issue for all electricity generators and research is ongoing, \cite{Barlow15,DREW2017a, cannon}, looking at the issues and solutions available to the sector to prevent curtailment, blackouts or large energy price fluctuations due to intermittent generation. It is proposed that predictable electricity generation may have a higher value than intermittent generation due to ability to plan grid balancing activities rather than having to react to sudden changes, \cite{shona2019}.\\ 

Slightly longer-term forecasts can be used for planning of marine operations where wave height or wind speed will limit the opportunities for undertaking work. Some operations may only take a few hours, while others a few days, and depending on the availability of vessels, equipment and personnel, they may need to be kept on stand-by waiting for appropriate weather windows. Improved accuracy for site condition forecasts can improve the planning of marine operations, reducing the stand-by time and therefore reducing costs. This may be more important when equipment within the WEC has failed and there may be safety risks associated with such operations being delayed. 

\subsubsection{Future perspectives and challenges}
A combination of historical and current environmental data can be used to determine migration routes for marine mammals and assess if there has been a long term impact on this by the presence of a wave energy array in a particular area. It is recognised that by creating a fishing exclusion zone within the array, it may displace marine life, even encouraging them to migrate towards the array. This, in itself is an interesting topic for which there is a lack of long term realistic data within the ocean energy sector, but possibilities of further research in the future are extensive.\\ 

Accurate site data, used for hindcast, real-time and forecasting is imperative to improve the design, optimisation, planning and long term operations of ocean energy technologies. As such, all methods for providing this data should be explored including how existing practices can be complimented and improved with satellite data. Satellite data could be a tool to help calibrating/validating hindcast wave resource numerical models. Indeed, although satellites would only be able to provide significant wave height and a wave length/period parameter at one particular time per day, this can be done over a wide area unlike what is available from conventional \textit{in situ} wave measurement (surface-following buoys, acoustic Doppler profilers).

\section{Satellite observations in the ocean}
\subsection{Satellite performance}

Different sensors are used within the field of satellite oceanography research. These sensors can be classified according to their range of measurement within the electromagnetic spectrum \cite{Preissner1978}. The performances of remote sensing instruments take place between certain wavelength (or frequency) ranges of the spectrum. Thus, optical sensors operate within the visible wavelength range, 0.4 to 0.7 $\mu$m  and the near infrared range between 0.7 $\mu$m and 1.1 $\mu$m.  Infrared (including thermal) measure between 1.1$\mu$m and 1 mm. Finally microwave sensors use wavelengths between 1 mm and 1 m (300GHz – 300 MHz). Atmospheric attenuation due to clouds, rain, snow, or water column varies according to the relative size of the attenuating features and the selected wavelength. Thus, optical and infrared signals are more affected by the meteorology than microwave signals which are much longer in wavelength and therefore scatter much less. Longer wavelength microwaves are typically therefore not altered by clouds, but can be disturbed by rain, especially at the shorter wavelengths used by many satellite radar systems. Sensors can be designed as multispectral, while others operate at specific wavelength. Multispectral instruments can operate at different wave bands (referred to as ``bands" as shorthand).\\

Another classification according to instrument performance is passive and active operation. Passive sensors are signal receivers, but do not transmit any electromagnetic pulse. These sensors can measure radiance, thermal emission,  and scattered energy that originates from some other source, such as the Sun. On the other hand, active sensors illuminate the planet surface with an electromagnetic pulse and measure the 'backscattered' signal of that pulse after it has scattered from features on the surface \cite[chapters 1]{Woodhouse2005}. Optical and infrared sensors are passive instruments, while microwave sensors are divided between passive and active instruments (radar).\\

Along with passive and active designation, sensors can also be classified as sounders or imagers. Sounders, also known as atmospheric sounders, measure vertical distributions of atmospheric parameters, such as temperature, humidity, precipitation and chemical composition. Infrared sensors are the most common atmospheric sounders, although microwave sounders are also in use. On the other hand, imagers measure two dimensional properties in a plane parallel to the surface. Optical, microwave, and infrared sensors can be used as imagers. Currently all the Earth Observation (EO) active radar systems are  monostatic, meaning that the receiver and transmitter are co-located on the same instrument. In the case where receiver and transmitter are different devices, these compose a bistatic system. Only the Global Navigation Satellite System Reflection (GNSS-R) and Refraction techniques use a bistatic system.\\

\begin{figure}
\begin{center}
    \includegraphics[width=\textwidth]{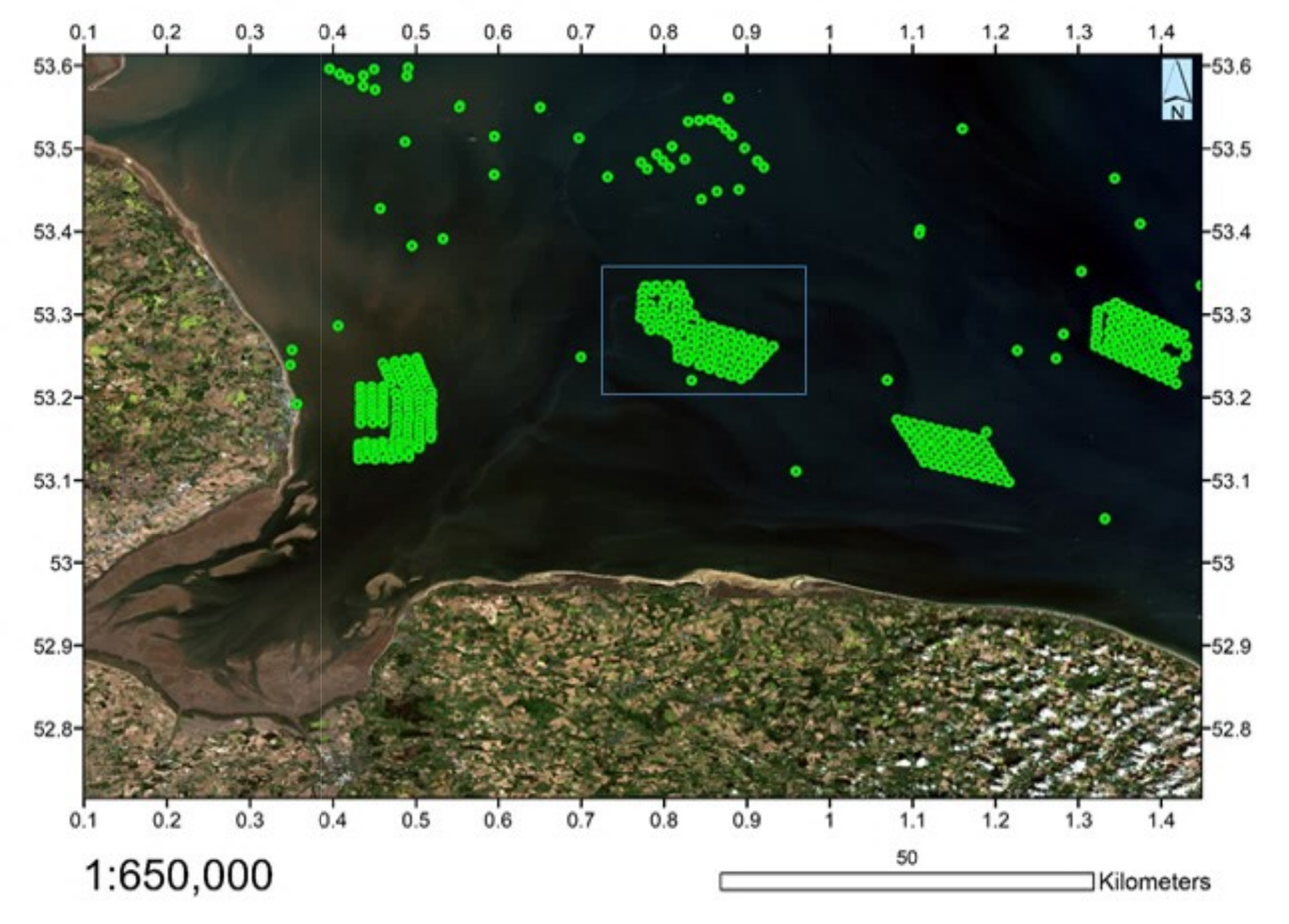}
    \caption{Synoptic view of offshore wind farms in the Humber region, southern North Sea. Four operational wind farms can be seen, with a further site under construction.  The background of the image is formed from merged Sentinel-2 red, green and blue visible bands at a pixel size of 10 m. An overlay of green symbols shows reflective objects in the scene detected with Sentinel-1 SAR. Both satellite products were collected within 24 hours (6th May 2020). Image files downloaded from ESA/EU \cite{copernicus} and processed with SNAP.}
    \label{fig:rabaneda1}
\end{center}
\end{figure}

\begin{figure}
\begin{center}
\includegraphics[width=\textwidth]{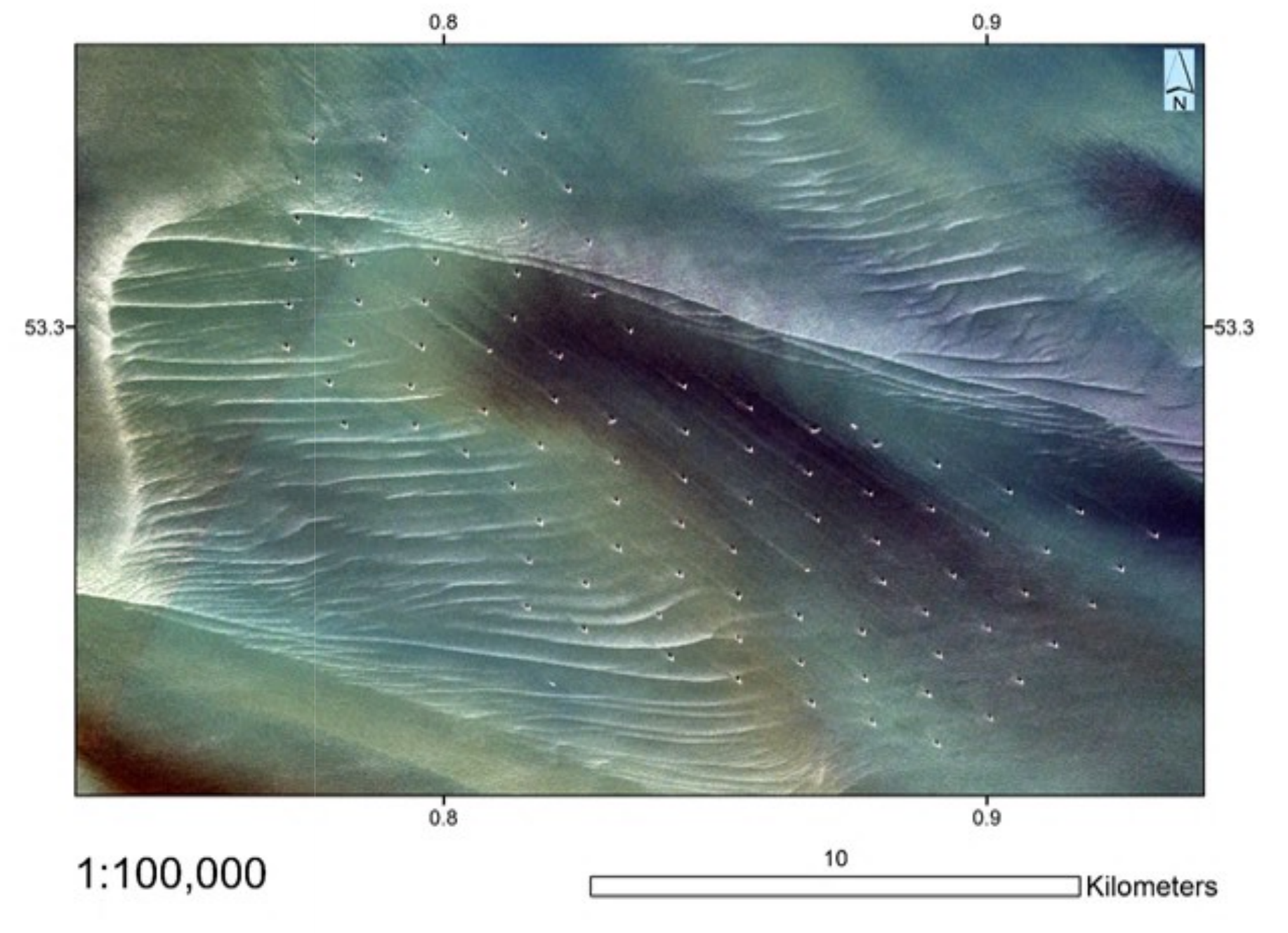}
\caption{Medium-resolution Sentinel-2 view of 580 MW Race Bank offshore wind farm showing details of underlying seabed bathymetry during a period of calm weather and transparent water. 91 wind turbines are visible within the farm, with some in shallow water (below 15 m) located on rolling sand wave structures, and others in deep water (above 20 m) where seabed details are not visible.}
\label{fig:rabaneda2}
\end{center}
\end{figure}

\begin{figure}
\begin{center}
    \includegraphics[width=\textwidth]{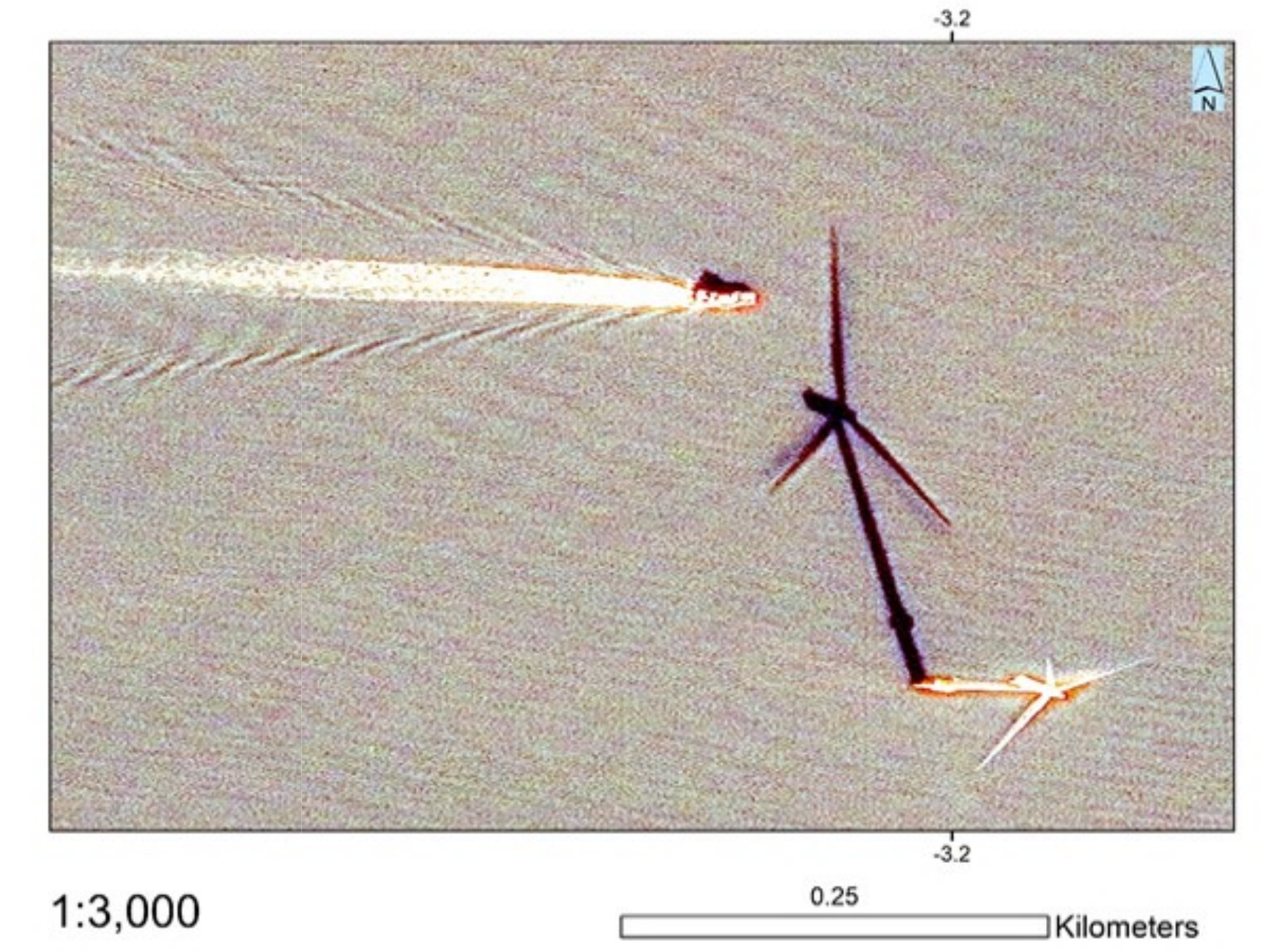}
    \caption{High-resolution Pleiades image with 0.5 m pixel size showing details of an individual wind turbine and passing crew transfer vessel.}
    \label{fig:my_label}
\end{center}
\end{figure}

Satellite on-board sensors can measure physical and environmental aspects of microscale ($\leq$ 1 km) and mesoscale (2 to 1,000 km) dynamics \cite[chapter 3]{Robinson2010}. However, the capability of a sensor to measure at different scales will rely on its spatial resolution, revisit cycle, orbit, and swath width (i.e. the area imaged on the surface as a satellite revolves around the Earth is called swath). Examples of different radar and optical satellite images are shown in Figures \ref{fig:rabaneda1}, \ref{fig:rabaneda2} and \ref{fig:my_label}. \\

Instrument performance is briefly introduced here for those instruments classified as imagers only. Then, passive imagers (optical, near-infrared, and radiometer instruments) are considered. Included active imagers are altimeters, scatterometers and Synthetic Aperture Radar (SAR). Finally, GNSS-R is also explained even though, strictly speaking, GNSS-R cannot be considered an imager. A comparison of general instrument characteristics is shown in \ref{table:sat_info}. Values and applications in \ref{table:sat_info} are meant to be indicative since they represent ranges of actual characteristics or applications. The reader is alerted to the fact that satellite remote sensing is a very active field of research, with new applications and enhancements being developed frequently, so that it is always wise to keep up to date with developments in this field.\\

All passive imagers follow a similar performance. They measure an electromagnetic signal emitted or scattered by the Earth surface, but do not illuminate the planet surface themselves.  Thus, passive instruments need an external electromagnetic source which will illuminate the planet surface --- this source is usually the Sun. This means all passive instruments typically follow a sun-synchronous orbit because they can only take measurements during daylight. The sun-synchronous orbit allows the sensor to maintain consistency of lighting conditions, since the local time below the satellite always remains the same. This is true for optical, near-infrared, multispectral, radiometers, and any other passive instrument. The difference between them is their design to work under different wavelengths or frequencies of the electromagnetic spectrum. Visible and near-infrared spectrum ranges were detailed above unlike radiometers, a term more often used to describe passive microwave instruments.\\

Active instruments send a pulse of electromagnetic radiation to planet surface and measure the backscatter in terms of time, power and polarisation. From simpler to more complex performance, actives instruments are altimeters, scatterometers and synthetic aperture radar (SAR). Altimeters send a pulse to the nadir, i.e. perpendicular to the planet surface with incidence angle equal to 0$^\circ$. Hence, altimeter swaths are narrow in comparison with others instruments. Since the speed of the pulse and height of the orbit above some terrestrial reference surface (datum) are known, the planet surface profile can be calculated by measuring the time it takes for the scattered signal to return to the sensor. The backscatter power is also measured for ocean wind speed applications. Scatterometers work similarly, but now the incidence angle is not null and backscatter time is used to locate the signal across a swath, rather than as an indicator of height. By measuring almost simultaneously a point with three different view angles it is possible to derive an estimate of both the wind direction and the wind speed. These instruments are also known as windscatterometers. SAR operation is similar to scatterometers but the spatial resolution is highly improved by the use `aperture synthesis', which involves the recording of the full backscatter signal of multiple echo returns by a single low-resolution antennas in multiple, sequential locations \cite[chapters 10]{Woodhouse2005}. \\

Finally, the global navigation satellite system reflectometry (GNSS-R) technique which, while a passive receiver only, takes advantage of signals transmitted in the microwave region by navigation satellites that hit the Earth's surface and are scattered back into space. It is not considered an imager \textit{per se} because its receivers can only measure points or lines of opportunity, only in one dimension, but these can be combined to create 2D images.  GNSS satellites themselves are only transmitters, such as American GPS, European Galileo, Russian GLONASS, or Chinese BeiDou missions. Therefore receivers are necessary to measure the scattered energy in a bistatic configuration (so called because the transmitter and received are in different locations); these can be on-board satellites or on Unpiloted Aerial Vehicles. The GNSS-R technique is not as mature as other active instruments, but it is not limited by the revisit cycle.\\

\begin{landscape}
\begin{table}[h!]
\begin{center}
\begin{tabular}{ |l|l|l|l|l|l|l| } 
\hline
\centering
\textbf{Instrument} & \textbf{Orbit} & \textbf{Revisit cycle} & \textbf{Swath width} & \multicolumn{1}{p{2cm}|}{\textbf{Spatial resolution}} & \multicolumn{1}{p{3cm}|}{\textbf{Raw measurement}} & \multicolumn{1}{p{3.5cm}|}{\textbf{Application}}\\
\hline
\hline
Optical & \multicolumn{1}{p{2cm}|}{Sun-synchronous, Polar} & 2 – 16 days & \multicolumn{1}{p{2cm}|}{35 – 400 km} & \multicolumn{1}{p{2cm}|}{$<$1 - 90 m} & Radiance & \multicolumn{1}{p{3.5cm}|}{Multi-purpose imagery (IWR), Ocean color/biology (OCB)}\\
\hline
Near-Infrared & \multicolumn{1}{p{2cm}|}{Sun-synchronous, Polar} & 2 – 16 days & \multicolumn{1}{p{2cm}|}{35 – 400 km }& \multicolumn{1}{p{2cm}|}{$<$1 - 90 m }& Radiance & \multicolumn{1}{p{3.5cm}|}{Multi-purpose imagery (IWR), Ocean color/biology (OCB)}\\
\hline
Radiometer & \multicolumn{1}{p{2cm}|}{Sun-synchronous, Polar} & 2 – 16 days & \multicolumn{1}{p{2cm}|}{1,000 – 1,800 km} & \multicolumn{1}{p{2cm}|}{25 – 50 km} &  \multicolumn{1}{p{3cm}|}{Brightness temperature} & \multicolumn{1}{p{3.5cm}|}{Ocean salinity (OS), Ocean surface winds (OSW), Sea surface temperature (SST)}\\
\hline
Altimeter & \multicolumn{1}{p{2cm}|}{Sun-synchronous, Polar} & 2 – 16 days & \multicolumn{1}{p{2cm}|}{~200 km} & \multicolumn{1}{p{2cm}|}{10 – 15 km} &  \multicolumn{1}{p{3cm}|}{Backscatter (time, power and polarisation)} & \multicolumn{1}{p{3.5cm}|}{Ocean surface winds (OSW), Ocean topography/currents (OTC), Ocean waves (OW)}\\
\hline
Scatterometer & \multicolumn{1}{p{2cm}|}{Sun-synchronous, Polar} & 2 – 16 days & \multicolumn{1}{p{2cm}|}{500 -1,800 km} & \multicolumn{1}{p{2cm}|}{12.5 – 50 km} &  \multicolumn{1}{p{3cm}|}{Backscatter (time and power)} & \multicolumn{1}{p{3.5cm}|}{Ocean surface winds (OSW)}\\
\hline
SAR & \multicolumn{1}{p{2cm}|}{Sun-synchronous, Polar} & 2 – 16 days & \multicolumn{1}{p{2cm}|}{20 – 400 km} & \multicolumn{1}{p{2cm}|}{10 – 2,000 m}  &  \multicolumn{1}{p{3cm}|}{Backscatter (time, power and polarisation)} &\multicolumn{1}{p{3.5cm}|}{Ocean surface winds (OSW), Ocean topography/currents (OTC), Ocean waves (OW)}\\
\hline
GNSS-R & \multicolumn{1}{p{2cm}|}{Geostationary, Medium and Low Earth Orbits} & \multicolumn{1}{p{3cm}|}{Possibility of 24h coverage} & \multicolumn{1}{p{3cm}|}{No swath, 1-D measurement of opportunity} & \multicolumn{1}{p{2cm}|}{No resolution, 1-D measurement of opportunity}  &  \multicolumn{1}{p{3cm}|}{Backscatter (power)} &  \multicolumn{1}{p{3.5cm}|}{Ocean surface winds (OSW), Ocean topography/currents (OTC), Ocean waves (OW)}\\
\hline
\end{tabular}
\end{center}
\caption{General characteristics of passive and active imagers, and GNSS-R.}
\label{table:sat_info}
\end{table}
\end{landscape}

\subsection{Satellite data processing}

The satellite based Earth Observation value chain pipeline is generally sub-divided into 3 categories; upstream, which includes the manufacturing and operations of satellites as well as their launch,  midstream, which includes government and commercial operators that sell or distribute EO data and finally downstream, which involves  the conversion of data into value added products. Despite the tremendous costs and innovation that occurs in the upstream category, to reach the stage of mature operational service provision it is the latter two categories which still pose the greatest challenges, but also the largest opportunity. Historically, all three categories of the EO value chain have been dominated by government and military organizations. The playbook consists of a government or military entity designing, building and launching a satellite (upstream), down--linking, processing and distributing the data freely on a platform (midstream) and providing some tools or algorithms for end users to extract specific value added information from the data themselves (downstream). \\

A good example of this public end-to-end approach is the European Union's Copernicus program. The program pulls together downstream data obtained by the  ESA, NASA, EUMETSAT and other government organizations environmental satellites, air and ground stations and sensors. The midstream data and information is processed and distributed free-of-charge to registered users on a centralized platform (Sentinel Hub, \cite{sentinelhub}). Finally, users are able to generate downstream value from the data through algorithmic tools provided either by Copernicus or developed by the users themselves. The most advantageous benefit from this type of end-to-end design is the availability of a wide range of data and information for free, and with an open licence to commercialise services on the back of this data (e.g. Sentinel 1 and 2, Figure \ref{fig:rabaneda1} and Figure \ref{fig:rabaneda2}).\\ 

While the public programs like Copernicus have been successful in providing a wide service to a variety of sectors, they are not optimised to fully meet the midstream and downstream demands of specific niche sectors like offshore renewable energy. Under these models, data is processed and delivered at too high of a latency, making real-time use of many data sets impossible. Additionally, users are required to have high technical capabilities in order to overcome the steep learning curve that comes with a Copernicus-enabled platform. Private companies like Orbital Micro Systems (OMS), \cite{oms}, are attempting to bridge gaps created by insufficient data latency, data storage and technical hurdles. Central to this approach is developing an infrastructure that improves the midstream and downstream categories in the EO pipeline. The two key elements of this infrastructure includes 1) utilizing more global ground stations for lower latency, and 2) utilizing cloud, Artificial Intelligence and Machine Learning technologies to develop a platform that is easy for users to generate value added information. Many companies, such as Amazon, have plans to dramatically increase their ground station networks, which will both reduce processing costs and enable real time extraction of data \cite{amazon}.

\subsubsection{Satellite data integration}

NASA and partners, including the US Weather Bureau, launched the first low orbit weather satellite, Tiros-1, in April 1960. Tiros-1 had cameras on board recording visible wavelength images of cloud patterns and weather systems. The excitement generated by this technological innovation was enormous as it marked a giant step change in the ability of the Weather Services to locate and track the development of key weather features over large areas, significantly enhancing forecasting capability.\\

Over the last 60 years, the influence of satellite EO on weather forecasting and climate monitoring has continued to grow enormously, moving well beyond images and strongly into the use of data. One of the most significant activities that capitalises on the value of satellite data is the process of data assimilation into computer--based numerical weather prediction models (NWP). Real--time satellite--generated data, along with measurement data from a range of other measurement technologies such as weather balloons, weather radar systems, aircraft measurements and weather buoys, are combined and assimilated into numerical weather prediction models. The evidence provided through these measurements makes it possible to nudge the computer simulation closer to reality, resulting in a more accurate starting point for the forecast. Since errors in the analysis can quickly lead to the growth of errors in forecasts (the butterfly effect), the importance of this data assimilation step is all too clear for forecast users, including the offshore renewable energy sector. There is also a strong reliance, of course, on state-of-the-art communication systems to facilitate the rapid dissemination of the measurement data, and satellites have a part to play there as well.\\

Due to the large spatial areas sampled by satellite sensors, including remote areas where it is currently difficult or impossible to make measurements with any other technology, satellite sensing now makes the biggest difference of all measurement platforms to improving the accuracy of the weather forecast. However, it is not just the forecast-dependent stages of renewable energy projects, Construction, Operations and Maintenance, and Power Production which benefit. Satellite data is also an important ingredient in the production of re-analysis datasets which are widely used in the sector for project planning and monitoring, risk assessment and site selection, for example in the simulation of inter-annual variability and economic viability. Re-analyses provide a multi-decade gridded climatological record of historical atmosphere and ocean variability on either a regional or global domain. The assimilation approach is similar to that used in NWP, except that the window of the  observations is not so time critical in this case. Also, the forecast model version used as a first guess is kept consistent over the duration of the re-analysis record. New re-analysis products are continually being produced, the purpose of each being different, for example long century scale records (e.g. \cite{laloyaux}), higher resolution within a specific regional area (e.g. UERRA, \cite{uerra}), close to real-time updates, \cite{hersbach}, or variables which are especially useful for particular user needs, such as 100m wind or wave parameters (e.g. \cite{borsche}, \cite{hersbach}).\\

\begin{figure}[h!]
    \centering
    \includegraphics{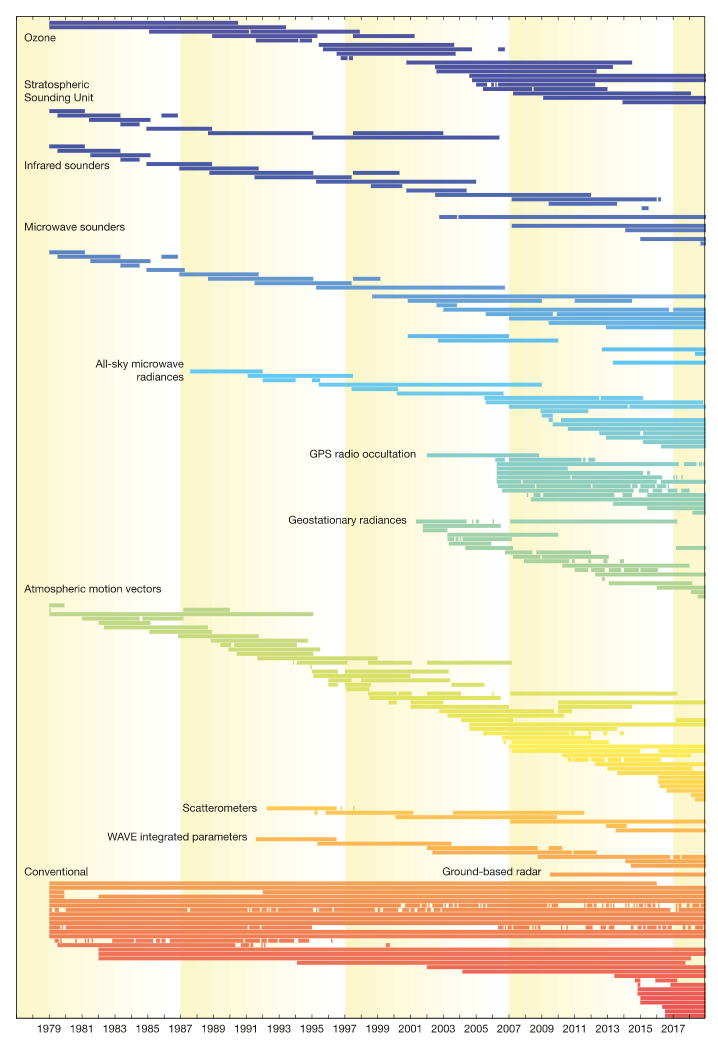}
    \caption{Data use in ERA5 since 1979. The horizontal bars represent the use of a satellite instrument or ground--based radar, or a source of conventional data, such as aircraft. weather stations, buoys, ships,  or radiosondes (image courtesy of Paul Poli).}
    \label{fig:dorling1}
\end{figure}

\begin{figure}[h!]
    \centering
    \includegraphics[width=\textwidth]{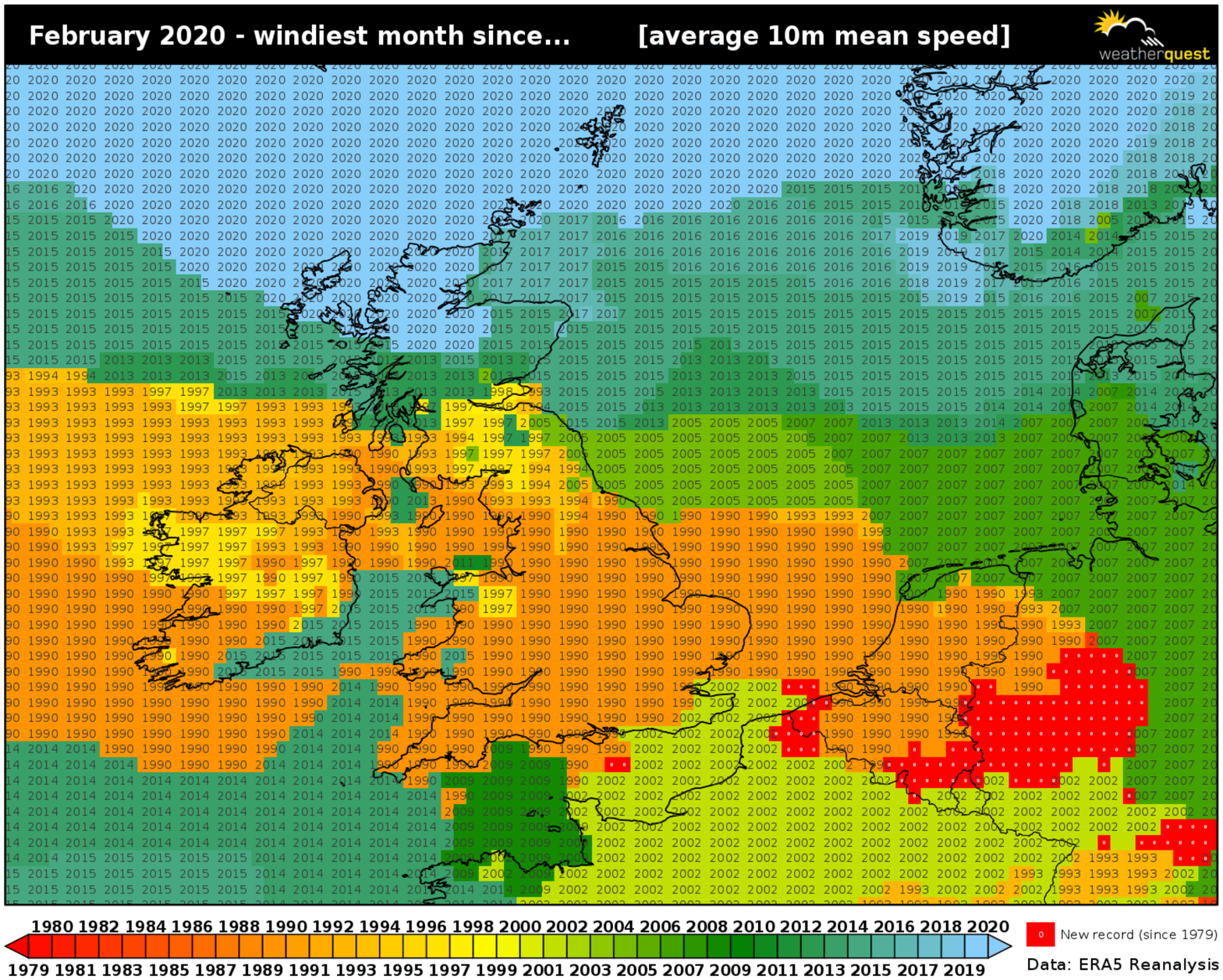}
    \caption{The previous year when monthly mean 10m wind speed exceeded that recorded in February 2020. Source: Weatherquest Ltd.}
    \label{fig:dorling2}
\end{figure}

\begin{figure}[h!]
    \centering
    \includegraphics[width=\textwidth]{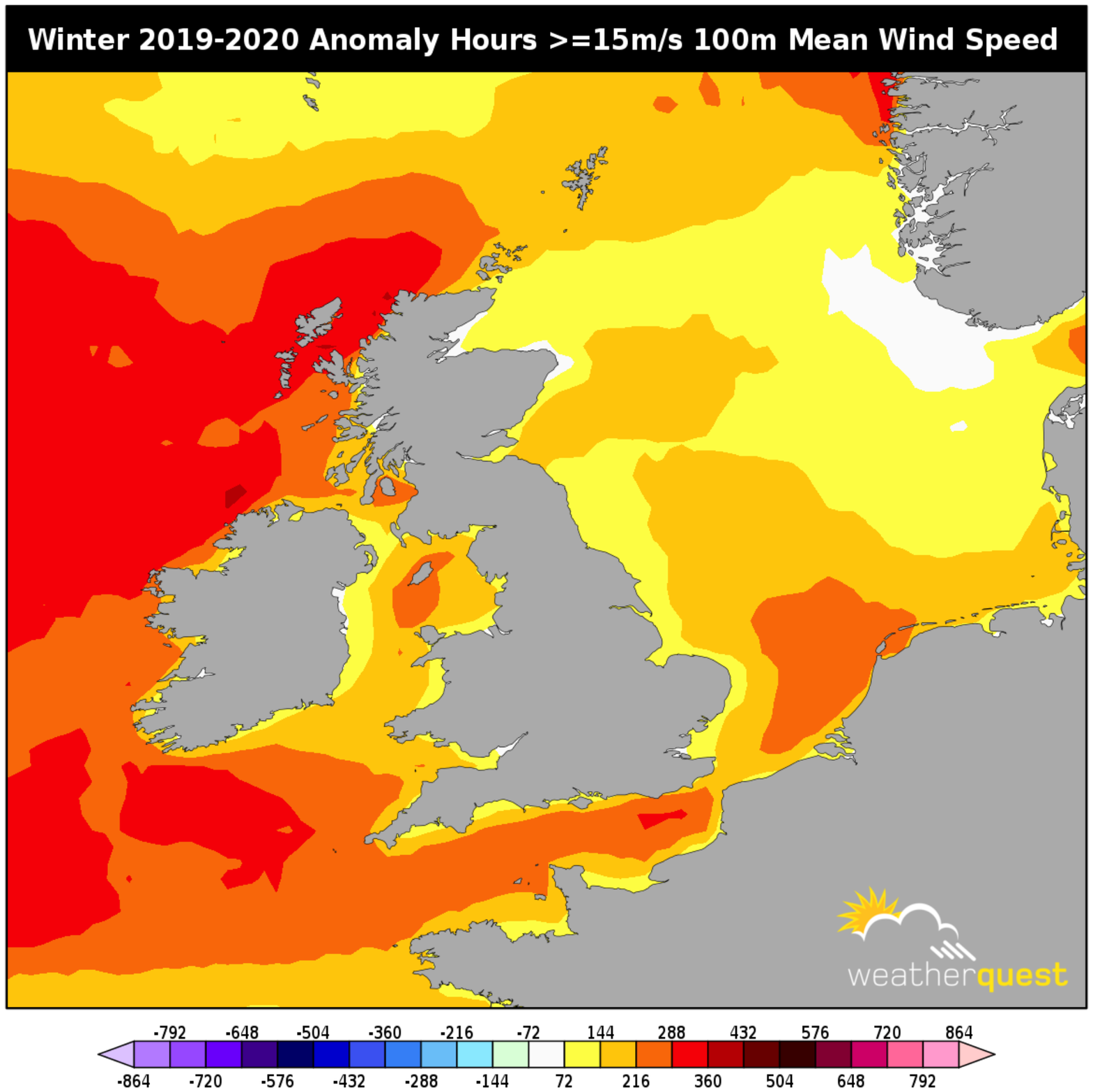}
    \caption{The number of hours with 100m mean wind speed $\geq 15$ m s-1 during Winter (December, January, February) 2019--2020 relative to the long-term ERA5 average. Source: Weatherquest Ltd.}
    \label{fig:dorling3}
\end{figure}

ERA5 is an example of the current state-of-science in re-analysis products, providing hourly estimates of atmospheric, land and oceanic variables at 31km resolution on a global and close to real-time basis, \cite{hersbach}. Figure \ref{fig:dorling1} shows the full range of observational data which have been assimilated into the ERA5 process since 1979, emphasizing the diverse and growing role of satellite data in climate monitoring. February 2020 was an especially windy month in North-west Europe, experiencing three named storms (Ciara, Dennis and Jorge); we can see in Figure \ref{fig:dorling2} how this compares with the long-term ERA5 records dating back to 1979. Meanwhile, taking the 2019-20 winter season as a whole, Figure \ref{fig:dorling3} shows how many hours saw more than $ 15 m s^{-1}$ wind speed at 100m height compared to the ERA5 long--term--average.\\

\subsection{Past, present and future satellite mission}

Satellite missions have been historically supported by governments through their space research organisations, such as NASA (National Aeronautics and Space Administration), ESA (European Space Agency), Roscosmos (Russian State Corporation for Space Activities), or CNSA (China National Space Administration). The first satellite missions were born during the Cold War, as a preparation step after the devastating consequences of World War II. Being able to dominate the space was the next step to prepare for any future international conflicts, and this started the so called ``space race", \cite{nasa1}. Since then, the global importance of satellites has grown to include the critical environmental component, and most of the public satellite missions nowadays focus on climate--related activities. The information provided by these is key for research towards sustainable development.\\    

\begin{figure}[h!]
    \centering
    \includegraphics[width=0.75\textwidth]{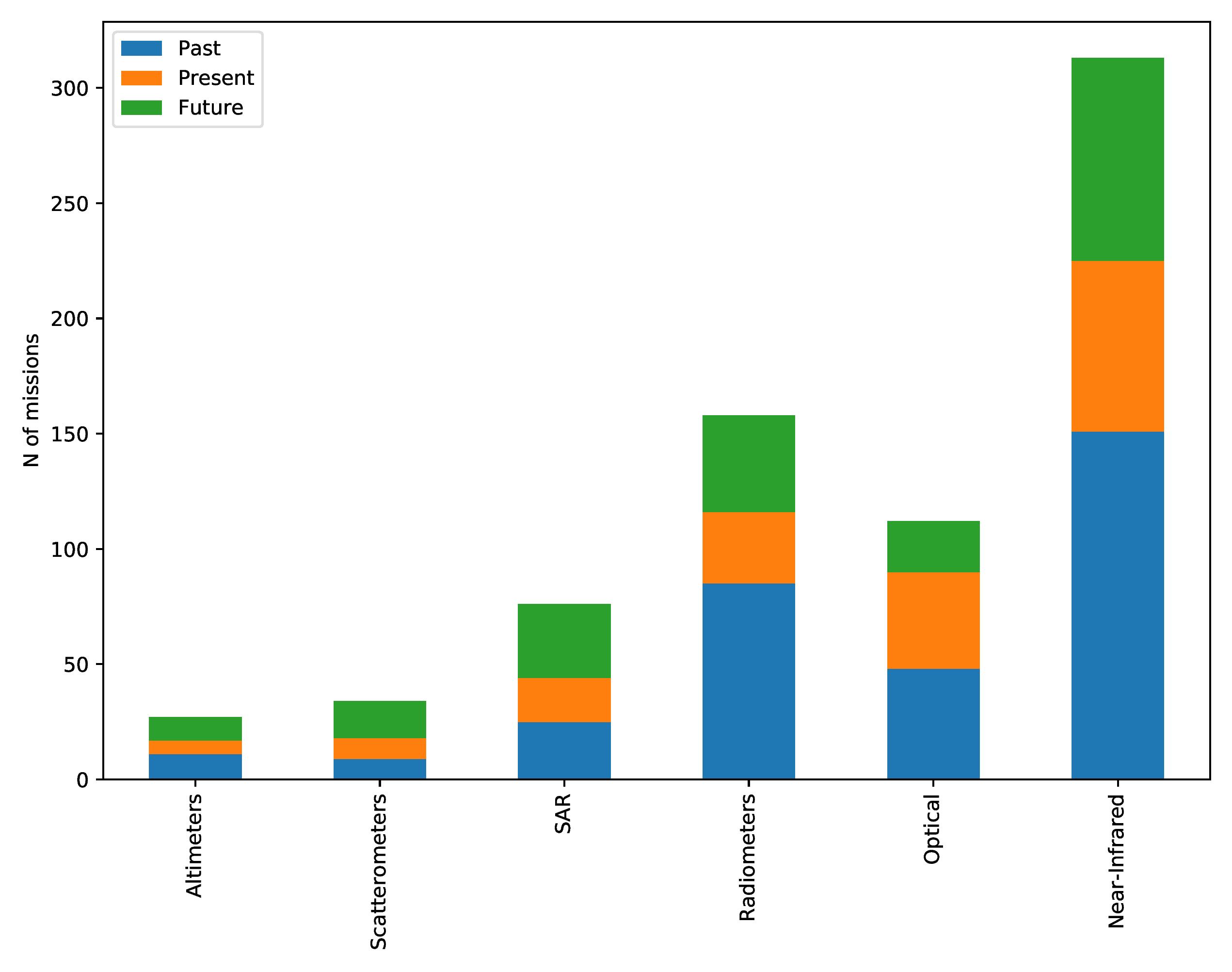}
    \caption{Number of missions by instrument. }
    \label{fig:by_instrument}
\end{figure}

The first satellite in orbit, Sputnik I, was successfully launched on October 4, 1957. This small satellite (58 cm in diameter) was the starting point of the space age. Since the launch of Tiros-1 in 1960, many missions have been completed. In those missions, satellites would carry one, or multiple, instruments on-board and those operated at different bands of the electromagnetic spectrum, from visible to microwave wavelengths. Thus, the amount of historical satellite  data is important. However, in the past much of the satellite data was not freely available to the general public. Research in EO and satellite instrumentation has enhanced the capabilities of those sensors in terms of resolution, swath, etc. Hence, a global picture of the satellite missions is necessary to understand the vast amount of data collected from satellites. The number of missions is too long to name each one. Instead, three different graphs are shown to illustrate the nature and amount of missions.  The source of information for these graphs was the CEOS database that collates information on all past EO missions, including current ones and those currently planned \cite{ceos_website}. The first bar plot, Figure \ref{fig:by_instrument}, represents the total number of missions by type of instrument. Sounders instruments were not included, only imagers. GNSS-R was not included either. Radiometers encompassed 3 different instruments: Earth radiation budget sounding radiometers, imaging multispectral radiometers (passive microwave), and multiple direction/polarisation radiometers. Near-infrared instruments were a blend of imaging multispectral radiometers (vis/IR) and hyperspectral imagers. In this case, near-infrared instruments use bands of the  spectrum at near-infrared wavelengths. Future missions included those with ``approved", ``being developed", ``proposed" and ``prototype" status.\\

\begin{figure}[htp]
    \centering
    \includegraphics[width=10cm]{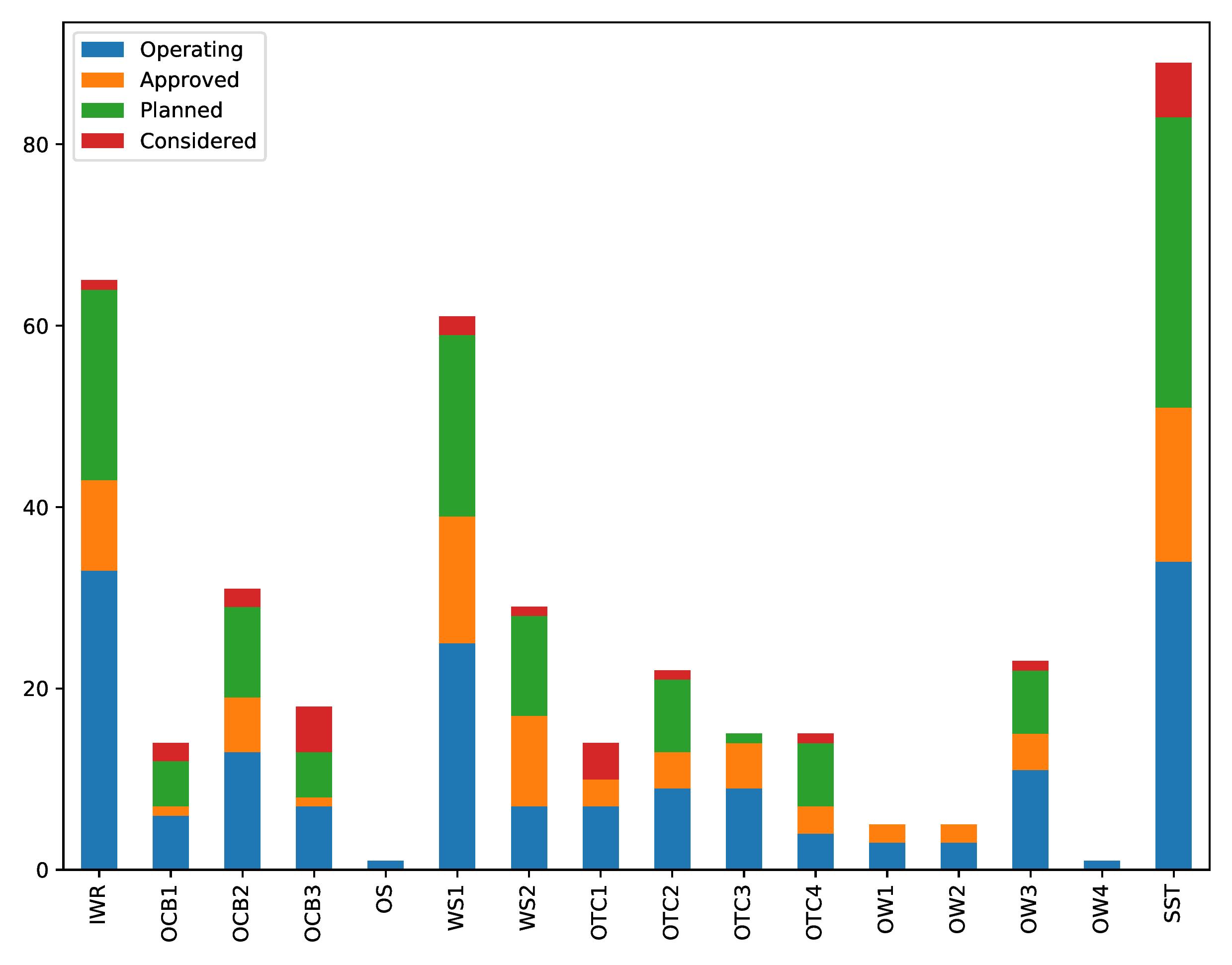}
    \caption{Number of missions by detailed measurements. Completed missions not included. From year 2000 to 2037. Acronyms meaning in Table \ref{table:acronyms}. }
    \label{fig:by_measurement}
\end{figure}

As shown in Figure \ref{fig:by_instrument}, near-infrared instruments have been used in more missions than any other. Radiometers are the second instrument in use. In meteorological applications, the near-infrared and radiometers are the two main satellite instruments used as input to Numerical Weather Prediction (NWP). Except for optical instruments, past missions outnumbered present missions. The same trend was observed between past and future missions, excepting for scatterometers and SAR. In the next graph, Figure \ref{fig:by_measurement}, past missions (i.e. completed), were not included. Only those with ``operating", ``approved", ``planned", and ``considered" status were included. Instead of arranging missions by type of instrument, in Figure \ref{fig:by_measurement} missions were arranged by the detailed measurement over the ocean only. Acronyms for the different types of measurement are detailed in Table \ref{table:acronyms}.\\

\begin{table}[h!]
\begin{center}
\begin{tabular}{ |p{1.7cm}|p{6.1cm}|p{4.8cm}| } 
\hline
\centering
\textbf{Acronym} & \textbf{Detailed measurement} & \textbf{Type} \\
\hline
\hline
IWR & Ocean imagery and water leaving radiance &Multi-purpose imagery (IWR)  \\
\hline
OCB1 & Colour dissolve organic matter &  Ocean color/biology (OCB)\\
OCB2 & Ocean chlorophyll concentration & Ocean color/biology (OCB) \\
OCB3 & Ocean suspended sediment concentration & Ocean color/biology (OCB) \\
\hline
OS & Ocean salinity & Ocean salinity (OS)\\
\hline
WS1 & Wind speed over the sea surface & Ocean surface winds (OSW) \\
WS2 & Wind vector over the sea surface & Ocean surface winds (OSW) \\
\hline
OTC1 & Bathymetry & Ocean topography/currents (OTC) \\
OTC2 & Ocean dynamic topography & Ocean topography/currents (OTC) \\
OTC3 & Ocean surface currents & Ocean topography/currents (OTC) \\
OTC4 & Sea level & Ocean topography/currents (OTC) \\
\hline
OW1 & Main wave direction & Ocean waves (OW) \\
OW2 & Main wave period & Ocean waves (OW) \\
OW3 & Significant wave height & Ocean waves (OW) \\
OW4 & Wave direction energy frequency spectrum & Ocean waves (OW) \\
\hline
SST & Sea surface temperature & Sea surface temperature (SST)\\
\hline
\end{tabular}
\end{center}
\caption{Types of measurements and their acronyms.}
\label{table:acronyms}
\end{table}

The most monitored parameters from satellites are sea surface temperature (SST) and wind speed over the sea surface (WS1). This was the expected because near-infrared imagers and radiometers can be classified as multi-purpose imagers, and many can measure SST and wind speed. Ocean colour biology (OCB) measurements are mainly taken by high resolution optical imagers, which can also be classified as IWR, see Table \ref{table:acronyms} for description. Ocean Waves (OW) and Ocean Topography/Currents (OTC) measurements are mainly measured by active microwave instruments: altimeters, scatterometers and SAR. Hence the number of missions for OW and OTC was low in Figure \ref{fig:by_instrument} for the passive instruments. Only one mission was found for ocean salinity, ESA's microwave radiometer for Soil Moisture and Ocean Salinity (SMOS). More information about this is provided in the section \ref{sec:phys}. \\

\begin{figure}[htp]
    \centering
    \includegraphics[width=10cm]{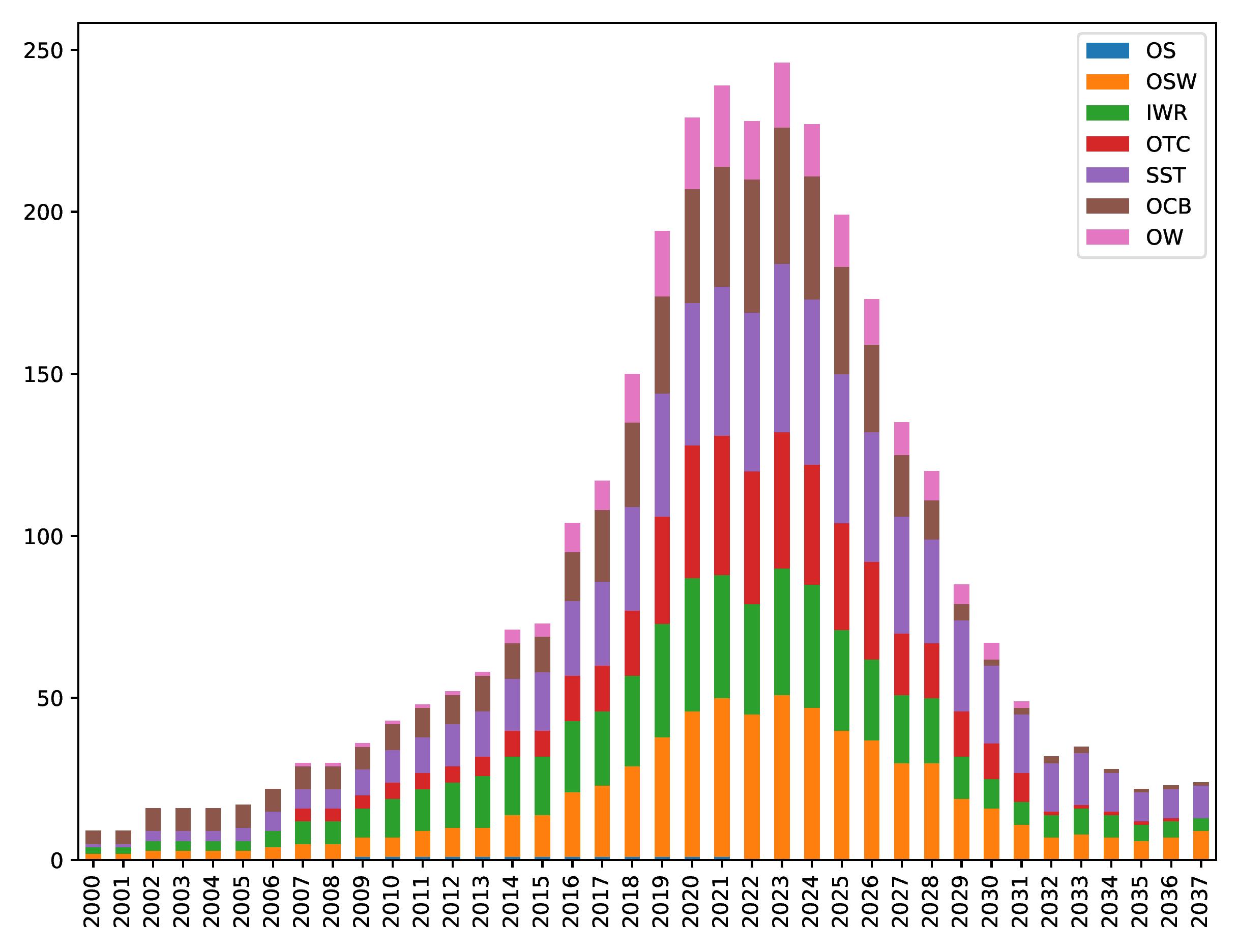}
    \caption{Number of operative missions by year and type of measurement. Completed missions not included. Legend acronyms in Table \ref{table:acronyms} }
    \label{fig:by_year}
\end{figure}

The last graph, Figure \ref{fig:by_year}, classifies the number of missions by years in operation and type of measurement. Completed mission are not included. Compared with Figure \ref{fig:by_instrument}, where all past missions are included, there are slightly fewer missions in the past than in the future and present together. This is also reflected in Figure \ref{fig:by_year} where the peak is expected to be located between the years 2021 and 2023. Thus, the exclusion of completed, i.e. not currently in operation, missions did not impact the numbers significantly. In the last decade, 2010 to 2020, the number of missions increased exponentially. On the other hand, between 2024 and 2031 a drastic decreased would be expected as planned missions are replaced by proposed missions in the database. However, more new missions are likely to be proposed in the coming years, filling the pipeline of planned missions. Proposal for new missions often rely on yet-to-be-agreed funding streams and/or the success of previous missions. In terms of measurement type, SST and OSW have the most missions coming in the next few years, followed by IWR, OTC and OCB. Only a few missions include observations of OW, and OS was not represented.\\

Missions are designated to a type of measurement according to the design specification for the instruments on board. However, due to improvements of algorithms for retrievals and advances in the relevant understanding of instrument measurements, some instruments are able to measure parameters for which they were not specifically designed. For example, Sentinel-2 carries a high resolution optical imager which was planned around making surface water reflectance measurements only (in the ocean science context); but it has now been demonstrated that OCB can also be measured \cite{Huizeng2017}. As another example, Sentinel-1 is a SAR operating in different modes and although it has a Wave Mode, there was no expectation that the Scan SAR modes (wide swath) would be able to infer wave properties, however, it has recently been demonstrated as possible \cite{Weizeng2016}.\\

\section{Satellite data for offshore renewable energy: synergies and innovation opportunities}

\subsection{Satellite data to reduce uncertainties for offshore wind farm operation}

\begin{figure}[!ht] 
\centering
\includegraphics[width=\textwidth] {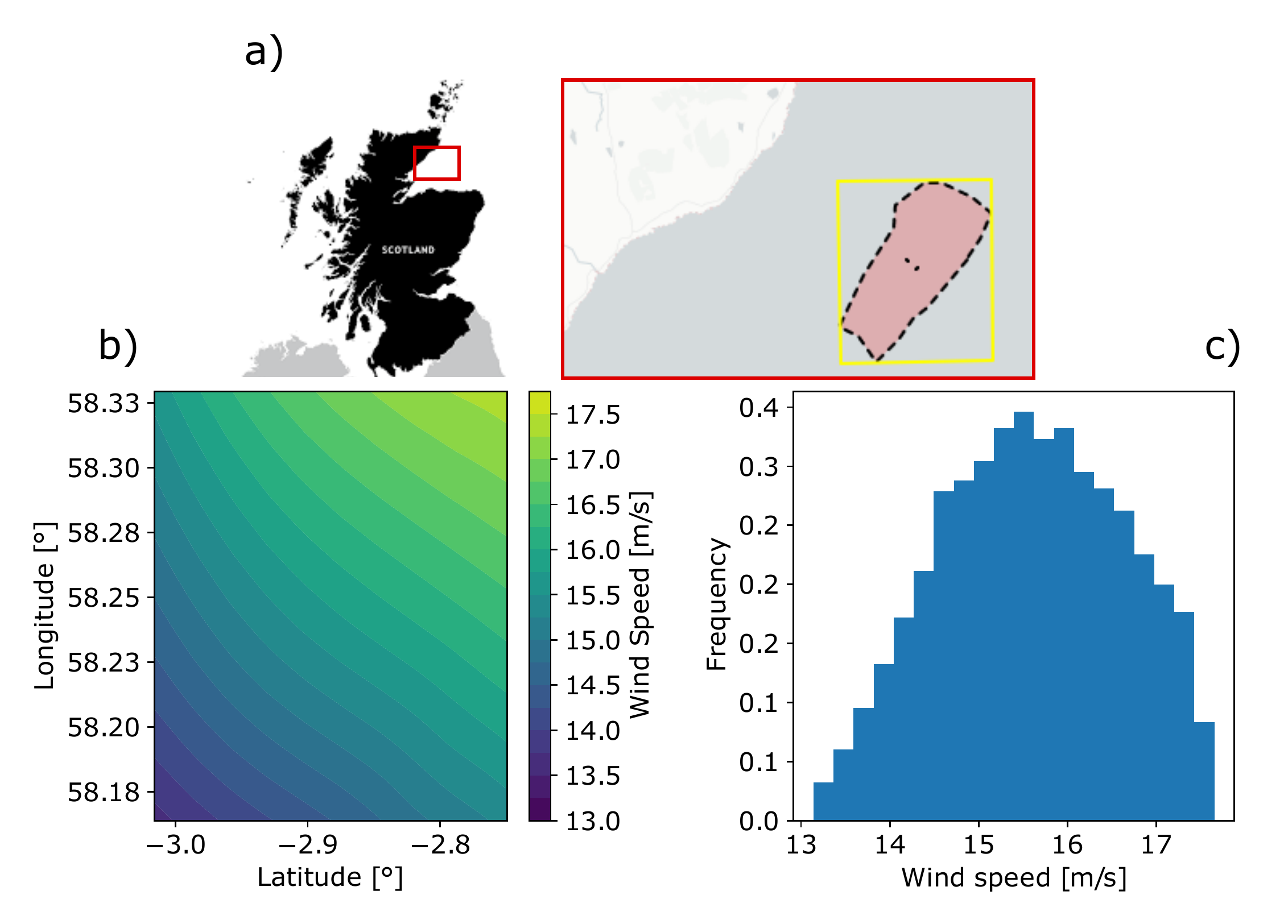}
\caption{Example of spatial changes in wind speed within the area covered by an offshore wind farm. a) Wind farm location, off Scottish shores.
b) Wind map for the area containing the wind farm presented in panel a). The wind field was extracted from  maps for wind speed retrieved from
SAR imagery available from DTU Wind Energy, \cite{dtu}. c) Normalized frequency distribution of the wind
speed value observed in b), \cite{zen2020}.}
\label{fig:uncertainty}
\end{figure}

Two types of weather--related uncertainties impact the management of offshore wind farms. One related to weather conditions and the difficulty of producing precise weather forecasts. The other is the uncertain impact of weather on operations, such as the difference between the estimated and actual energy production or accessability for maintenance, see Figures \ref{fig:uncertainty} and \ref{fig:windpowerforecast}.
These two factors interact motivating the development of application--specific solutions.
To quantify time--related uncertainty in global weather models, \cite{cannon} perturbed the forecast to assess the temporal scales for which a weather forecast can be considered more reliable than long-term climatological statistics. \cite{browell} follows a similar line, where the difference in terms of cost from adopting deterministic or probabilistic approaches to estimate access windows for O\&M works is quantified. Results showed that $4\%$ of the available weather--windows are not used by wind farm O\&M, and that the use of a probabilistic approach could reduce such percentage with minimal cost impact. Furthermore, they showed that by adopting a cost--loss decision process incorporating  the probability of successfully performing maintenance and the cost of the energy, overall economic operation  could be improved, i.e. higher cost of lost energy capture will increase the incentive to restore an unavailable turbine. \\

Coarse spatial resolution of wind and wave data is also a significant factor in these uncertainties. Data of wind speed and sea state available from global and national forecast providers present a coarse resolution with information distributed on a grid size that ranges from 1.5km (e.g. Met Office models for the UK) to 80km, resulting in having one value available for the entire wind farm. Local measurements used to inform and calibrate meteorological models are also sparse. As a result, one value for wind speed and wave height is assumed to be representative for the entire area and used by wind farm managers to direct operations. The high degree of uncertainty in local wind and wave conditions, possibly due to the interaction with the turbine, further increases the difficulties in choosing access windows that will allow safe access to offshore structures. Moving to a turbine-specific approach is desirable, especially for larger wind farms and those with complex bathymetry where conditions across the farm can vary significantly. In this regard, higher spatial and temporal resolution data from satellite missions could significantly improve the operation of wind farms.

\subsection{Satellite data for short--term power forecasting}

\begin{figure}[t] 
\centering
\includegraphics[width=0.7\textwidth] {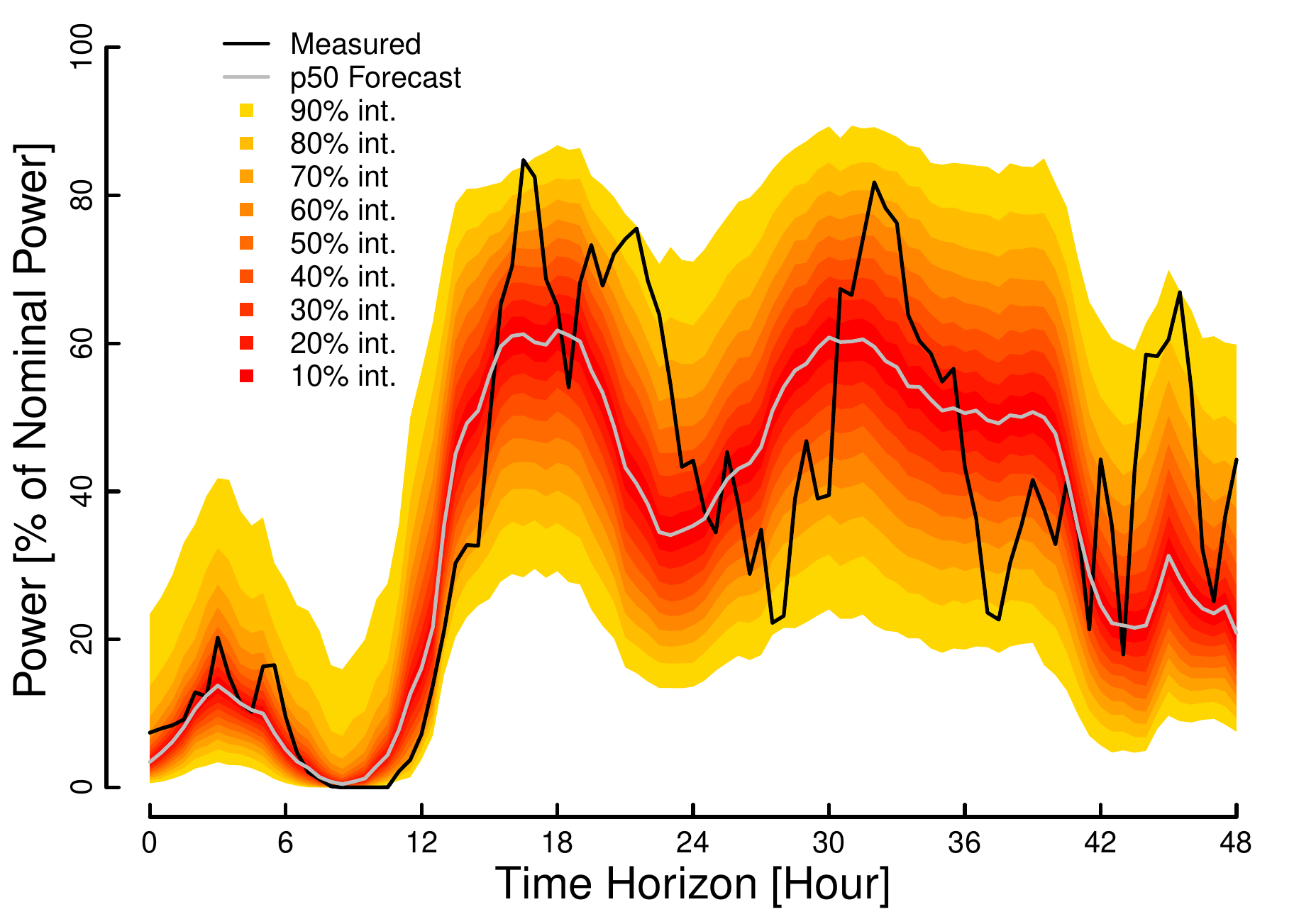}
\caption{An example of a probabilistic wind power forecast from \cite{Gilbert_2020} (under licence CC-BY-4.0). Uncertainty quantification, in this case via prediction intervals, provides valuable information to decision-markers seeking to maximise profits and/or manage risk. Reducing uncertainty, i.e. being able to make such forecasts `sharper' while remaining calibrated (unbiased at all probability levels), would increase their value.}
\label{fig:windpowerforecast}
\end{figure}

Power forecasting, the prediction of future energy demand and production, plays a relevant role in the operation of electricity systems where supply and demand must balance on a second-by-second basis. This task becomes more challenging as the penetration of weather dependent renewables and electrification of other energy vectors, notably heat and transport, increases. Short-term forecasts of wind power production are required for individual wind farms and aggregated by market participant or geographic region~\cite{Bessa2017}. They are used by generators to inform energy trading and electricity network operators to maintain a reliable and economic supply of electricity. Wind farm operators also utilize forecast to support asset management processes, such as maintenance access, particularly offshore. The same will be true of other ORE technologies if/when they reach installed capacities that necessitate short-term forecasting. An example of a wind power forecast is provided in Figure~\ref{fig:windpowerforecast}.\\

Short-term energy forecasts are typically produced by post-processing NWP to convert meteorological forecast in to energy-specific forecasts, such as the power output of a wind farm \cite{Sweeney2019}. Uncertainties and errors result from both the NWP process and weather-to-power conversion. Research is ongoing in both areas to improve forecast skill. Satellite data has been responsible for substantial improvement in NWP skill over the past 40 years and will likely be so in the future \cite{Eyre2019}. Innovations such as ESA's Atmospheric Dynamics Mission Aeolus (ADM-Aeolus)---the first satellite to be able to observe wind profiles at a global scale--- may yield significant improvements in wind field forecasts over seas and oceans, where observations are sparse today, and therefore improvements in offshore wind power forecasts too. A limitation of NWP is the latency introduced by data assimilation and the computational time required to run the models themselves. By the time forecast data is available, the most recent observation assimilated may be several hours old and even short-range NWP forecasts are subject to epistemic (or systematic) uncertainty. Therefore, very short-term horizons (minutes to hours-ahead) forecasters incorporate live measurements of power production and meteorological variables, and may not utilise NWP at all.\\ 

Satellite images have been used to produce intra-day (hours-ahead) solar power forecasts and bridge the gap between purely NWP-based predictions and local observations from sky cameras and measured power production \cite{Blanc2017}. Estimates of surface solar irradiance and cloud motion vectors are derived from these images, which, crucially, are available with latency of only minutes. These methods are likely transferable to the wind power forecasting but satellite observations of wind fields are not available at the requisite spatial or temporal resolution. Observations would be required within 10s of kilometers of a target wind farm with a refresh-rate of one-per-hour or higher. Similarly, very short-term sea state forecasting, used for offshore maintenance operations, could be improved if the sea state could be inferred from satellite data on similar scales.\\

Power ramps, large changes in power output over short time periods, are a consequence of highly concentrated renewable generation capacity, such as large solar or offshore wind farms \cite{DREW2017a}. Power ramps may be associated with synoptic-scale (or large-scale) weather features, such as passing frontal systems, or localised events such as convention or cloud formation/dissipation. The existence of synoptic-scale weather features is well predicted but their precise location at a given time may not be, where as convective process are a common source of error. The precise prediction of ramp timing, rate and duration is an example where improved (very) short-term forecasting would have high value and may be realised though use of satellite data.\\

\subsection{Satellite data for wave forecasting}
\label{sec:wavesat}

Accessibility of  devices for personnel from various classes of service operation vessels (SOV) and Crew Transfer Vessels (CTV), is key to the planning and execution of operations and maintenance of offshore developments \cite{evdokia1}. Accessibility is determined by numerous parameters, most importantly: sea state conditions \cite{evdokia2}, required vessel transit time; and vehicle type \cite{evdokia1}. The transfer and deployment of personnel and equipment to individual offshore locations is a significant cost factor in routine, and unplanned, maintenance and inspection. Sea state is the general conditions of the sea surface. In marine engineering, sea state is often characterized by the significant wave height \cite{evdokia3}. With operations and maintenance comprising $25\%$ of the overall cost of a wind farm, \cite{evdokia4}, $25\%$ for wave, and $15\%$ for tidal, \cite{report}, accurate spatial and temporal forecasting of sea-state is critical to reduction of the LCOE.\\

Forecasting of sea-state parameters, including significant wave height and period, is traditionally performed through numerical models following a set of physical rules to simulate the creation and propagation process of waves within an area \cite{evdokia5}. Forecasts are provided at various spatial resolutions, and with different lead times, with varying degrees of accuracy and computational cost. A number of open-source physical forecasting models are widely available, and are used within the offshore industry for planning operation and maintenance \cite{evdokia6}. Some of the most widely used models are the: WAve Model, WAM \cite{evdokia5}; Simulating WAves Nearshore, SWAN \cite{evdokia7}; and WaveWatch–III, \cite{evdokia8}. These models provide both hindcast and forecasts of ocean waves at variable resolution. However, when compared with data from floating buoys or wave radars the accuracy of physical forecasts depends on several parameters, such as: forecast lead-time; expected significant wave height; correlation with meteorological conditions, including wind speed, direction and fetch; and bathymetric and environmental flow conditions \cite{evdokia6}. Moreover, the spatial variability is limited to the grid size used by the respective forecasting model. However, smaller grid sizes require non-linear increases in computational cost \cite{evdokia9}, making them impractical for fine-resolution short-term, live forecasting. Currently, the operational model widely used for sea state forecasting in the UK is the Atlantic – European North West Shelf (Copernicus Marine Environmental Monitoring System -- North West Shelf Seas, CMEMS--NWS). Ocean wave analysis and forecast are provided on a regular grid at 0.016 degrees sourced by CMEMS \cite{evdokia10}. Physical models are useful for many applications, yet their spatial resolution is not adequate for maintenance planning in offshore wind farms, which cover range of area in $1-3$ km.\\

\begin{figure}[t] 
\centering
\includegraphics[width=\textwidth] {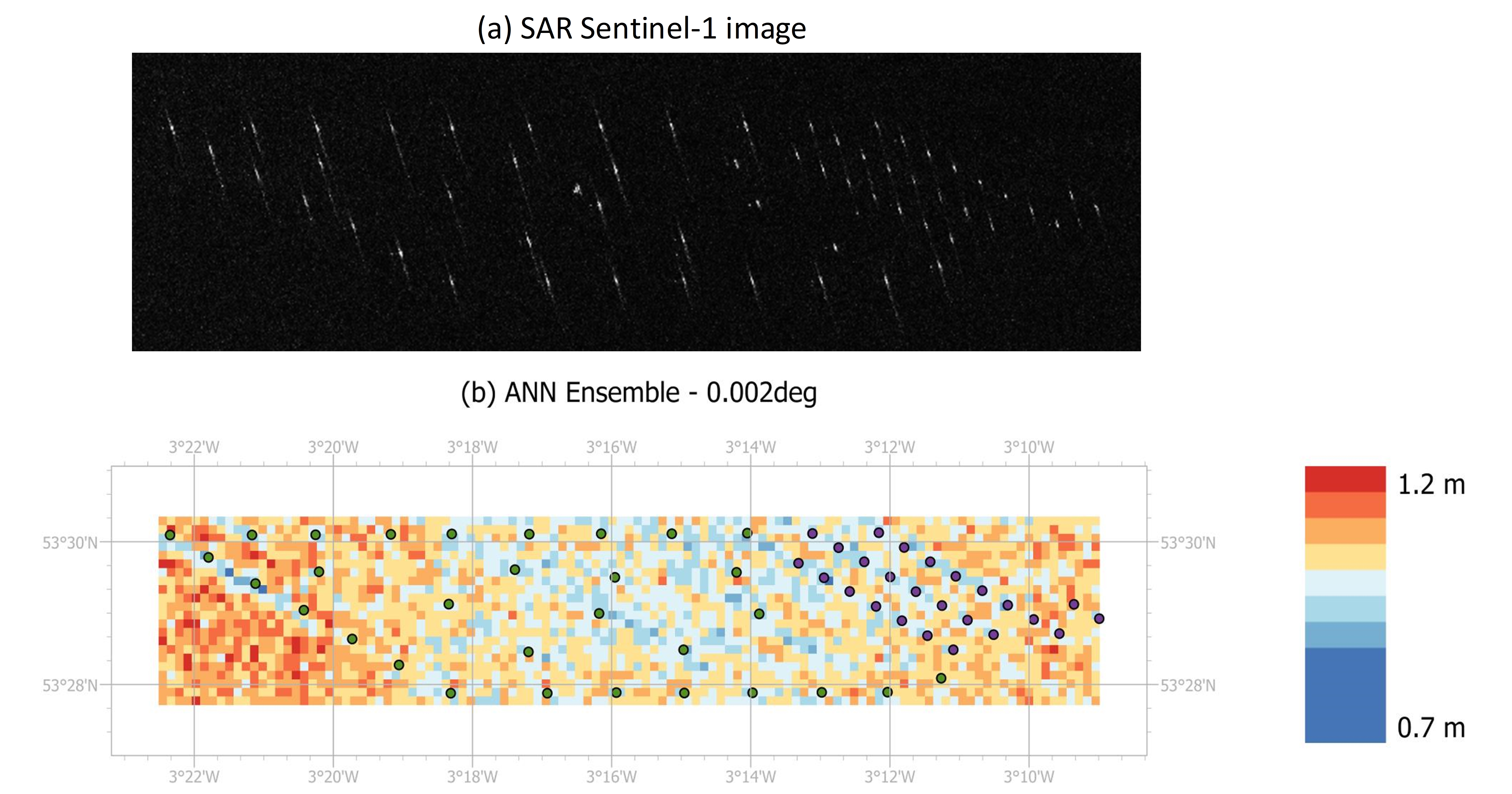}
\caption{Sea state conditions at 2/4/2019 at 06:30am. In coloured dots the positions of the Burbo Bank and Burbo Bank extension wind turbines are denoted. In (a) the SAR Sentinel-1 satellite image while in (b) the hindcasted SWH using an ANN ensemble model in 0.002 degrees resolution. Modified after \cite{evdokia14}.}
\label{fig:wave_forecasting}
\end{figure}

Satellite remote sensing in combination with machine learning can help overcome the problems faced by physical forecast models. In particular SAR sensors, such as the ESA Sentinel-1 platform, can provide images in a bi-daily resolution in best cases independent of meteorological conditions and cloud coverage \cite{evdokia11}. With recent advances in satellite mapping frequency, \cite{evdokia12}, global remote monitoring from satellites now provide the quantity of data needed to train and validate Artificial Neural Networks (ANN) correlating SAR images to direct and forecast measurements of sea-state. When trained against a point measurement (i.e. wave buoys and wave radars), ANN processed SAR images provide high spatial resolution of sea-state. Open source spatial  resolution of ESA Sentinel SAR images is down to 5 m$^2$, over two-orders of magnitude finer than open-source sea state models \cite{evdokia12} (although readers should note that practical speckle-reduced imagery will be nearer 15m resolution in practice). In Figure \ref{fig:wave_forecasting} an example of a SAR image and the resulting SWH hindcast for Burbo Bank wind farm is presented.\\

Using ANN’s, the significant wave height at native resolution of satellite images is available. The advantages of such a high fidelity hindcast from a SAR image are multiple, including: 1) finer resolution information than any other source of information available; 2) they are unaffected by cloud cover and meteorological conditions; and 3) available in normal time intervals (i.e. hours). Moreover, further work is needed to use ANN’s to integrate long-time series of SAR data and develop computationally efficient sea-state models below the resolution of numerical forecasts. Such nested models balance the need for fine-scale spatial resolution, accuracy, and forward temporal forecasting needed to optimize operations and maintenance, reducing the LCOE in offshore wind energy sector \cite{evdokia14}.\\


Contrary expectation, in this application better satellite image resolution would not improve the results of such a model. The resolution of Sentinel-1 SAR images is already high enough, causing a very noisy spatial distribution of significant wave height that forces the grouping of neighboring pixel values in order to get a usable result. To produce optimal results, resolutions similar to the wavelength of the hindcasting wave conditions are preferred. Optical images from other satellites cannot be used in similar ways with current deployments, since the number of usable images (low cloud cover, different lighting conditions per day) reduces the reliability of the results. However, introducing all-weather remote sensing technologies and improving radar satellite capabilities, their temporal resolution could vastly improve the model and extend its usability into live forecasting of wave height in offshore locations.\\

\subsection{Satellite data for tidal forecasting}





The issues here are very similar to those for the wave sector. Explicit to tidal stream is the need for prediction of extreme wave--current interaction events. This currently relies on a good quality 3-D regional fluid model (preferably two--way coupled wave--current systems) to provide predictions of the fluid velocity field and turbulence. Satellite data have substantial scope to support the development of tidal range schemes. In the first instance, satellite data can provide data to attest the accuracy of numerical models employed in the resource assessment of tidal energy. Satellite imagery can be used to evaluate whether coastal models capture inter-tidal and inundation processes, that are crucial for impact assessments of the schemes. In addition, data can be used to help identify changes in salinity and sedimentation processes using optical data, see section \ref{salinity}, and SAR and altimetry  techniques. Satellite data can be beneficial indirectly, for example, storm surges need to be predicted using satellite data that serves as input to forecast models. There are  opportunities for innovation in  employing satellite data to quantify the impact of large scale tidal energy developments. For recently developed schemes, such as in the Lake Sihwa tidal power station, \cite{sihwa}, satellite data before and after the construction can provide an insight to inundation and water quality changes due to the presence of the structure.\\

\subsection{Satellite data for environmental monitoring}
\label{sec:env}
Environmental monitoring is a key tool to minimize environmental degradation and pollution, disappearance of cultural heritage and landscape, or impacts on human health caused by human interventions \cite{sadler1999principles}. Environmental legislation, such as European Union Environmental Impact Assessment (EIA) Directive (85/337/EEC) and National Environmental Policy Act (NEPA) of 1969, has been improving over the years with multiple adjustments. Although EIA has strengthened its role in decision-making processes over the last 15 years, understanding anthropogenic impacts on the environment is still incomplete due to lack of adequate monitoring programmes \cite{jha201625} and the scarcity of ORE installations \cite{mendoza2019framework}.\\

To date, EIAs of ORE developments are mainly site--specific and device--specific \cite{mendoza2019framework},and for an effective environmental assessment, proper guidance is needed, as well as best practice examples, better public engagement, and a decision-making process based on evidence and multiple criteria, \cite{campos2019life,fischer2019editorial, sinclair2018implementing, fischer2019editorial}. There is still the need for much research to achieve a universal methodology for the EIAs \cite{mendoza2019framework}, which should also include new long-term climate change targets within operational and mitigation frameworks of future projects  \cite{fischer2019editorial}. It is important to consider climate change effects alongside environmental and physical changes, following ORE developments (Figure 17). Exploring how both of these pressures will change marine ecosystems requires a holistic approach to consider interactions between species and their environment. This type of assessment requires long-term data sets across multiple trophic levels, thus being able to utilize different sources of information is critical to support the EIA process and decision-making. In addition, pressures need to be set in the context of cumulative pressures from other marine industries and users in the area to aid the identification of potential frameworks for in-combination assessments across multiple sectors.\\ 

Given the cost and capacity related to data collection, science--based management must focus on the optimisation of the use of existing data and  evaluate the relevance of new data explicitly within a value-of-information framework \cite{burgess2018five}. Conventional EIA relying on direct sampling and site-specific analysis is a costly, time and resource consuming process that can offer a direct and precise impact analysis but limited to the sampling effort, area and time series \cite{patil2002comparison}. Thus, remote sensing offers a different set of data at a much extended temporal and spatial coverage, representing a relatively cheaper tool to achieve a geospatial cumulative EIA for large and long-term installations, but presenting some limitations related to image resolution and targets \cite{patil2002comparison}. Still, analysis of remote sensing data can be used as an effective tool in EIA studies to assist the understanding of complex interactive effects within the physical and biological environment, to support the understanding of cumulative effects and minimise uncertainty with future climate change to aid the evidence-based decision-making process \cite{patil2002comparison, moufaddal2005use}. Satellite data allow evaluation of the environment and monitoring of changes throughout different spatial and temporal scales, which can be used in combination with field observations and expertise knowledge. An example of this integrating technique is the spatial multi-criteria analysis implemented through geographic information systems (GIS), where the combination of multiple heterogeneous dataset (objective data and subjective values, e.g. key public concerns) into a geospatial analysis would help in visualizing, evaluating and structuring EIA for current and future scenarios \cite{gonzalez2018spatial, gonzalez2019designing}. \\

\subsubsection{Physical processes}
\label{sec:phys}

Organisms need to adapt to their habitat to be able to survive. Understanding the physical characteristics of a habitat and how these might change with climate and placement of ORE devices is a key area and requires an understanding of the physical processes, related to hydrology and geomorphology. Currently, physical habitat models (e.g. Scottish Shelf Model, \cite{shelfmodel}) are a reliable tool to investigate the impacts of tidal and wind energy on the environment. Extraction of energy from tidal and wind devices reduces the energy in the environment and causes local and far field hydrodynamic changes. Intertidal habitats can be affected by the variation of the principal tidal harmonics (sinusoidal components with amplitude and frequencies determined by local conditions, the sun and the moon gravitational forces), which causes temporal changes of the ebb-flood cycle \cite{de2018comparative}. Fish and top-predators rely on these repeated tidal patterns, but scientific evidence of the possible effects of this variation are lacking. The most known consequence of wind farms relate to the atmosphere: the “wake effect” (the wind speed reduction behind wind turbines) can induce a change of air temperature and sea surface pressure up to 15 km away from the wind farms \cite{hasager2013wind} causing local climates variations and cloudiness. \\

The foundation of every wind turbine can be seen as an obstacle in the sea and their aggregation within farms might lead to an impact on the horizontal and vertical currents of the sea circulation within a local spatial scale. Changes of currents intensity or directions might influence the sediment transportation, stratification and mixing rate of coastal and offshore waters, and consequently affect the primary productivity of these regions by altering the nutrients supply and light availability in the photic layer (surface part of the sea penetrated by sunlight). Few studies have included analyses of the impact on primary production, but their results differ among regions due to their unequal topographic and oceanographic conditions. In shallow waters ($<$ 25 m), it has been reported that the reduction of wind pressure causes an increase of net primary production due to a decrease of mixing events and a sediment load \cite{van2014predicting}. However, the effects on offshore, deeper, waters are not yet documented. Tidal turbines might also affect the primary productivity alongside the channels, however the consequences could appear at larger scales than the weaker effect, over hundreds of kilometres away from the turbines, causing the formation of new shelf banks, less turbidity and an increment of primary production with associated faunal ecosystem effect \cite{van2016potential}.\\

Hydrodynamic models have become an important tool for marine planning and management due to their higher resolution compared to reanalysis products (e.g Copernicus, \cite{copernicus}) and representation of OREs \cite{de2018comparative}. However, some of their limitations include computational costs and their uncertainty assessment, which is required to aid EIAs and support decision-making based on model outputs. Empirical data (e.g \textit{in situ} and remote sampling) then becomes necessary to evaluate model predictions related to physical characteristics, such as topography, stratification, currents, tides, temperature, salinity, sediment transport, water types and quality. Satellites have been largely adopted in oceanographic and ecological EIAs, although it is important to keep in mind that they capture most of the data coming only from the first layer of the water column. Today, many ocean colour products are available for long time series, due to different initiatives (e.g. Copernicus, \cite{copernicus}; ESA, \cite{esa}; NASA, \cite{nasa}) which elaborated, standardized and validated many variables (Figure 17A, B). Oceanographic features that are ecologically significant for the marine environment (phytoplankton fronts, temperature fronts, internal waves, upwelling and downwelling regions) can be derived from satellite images and their spatial distribution and temporal pattern should represent the key factors for EIAs. On occasion, bathymetric features on the seabed itself can be clearly resolved in medium resolution satellite images, and could be used to detect scour and wake effects on the surrounding environment. Moreover, understanding of the spatial and temporal scale and further \textit{in-situ} habitat data can assist the predictive physical models’ uncertainty and enhance sensitivity.\\

\subsubsection{Salinity and temperature}
\label{salinity}
The issues that ocean salinity and temperature \textit{in situ} measuring techniques present (e.g. low spatial resolution, poor coverage in coastal areas, or expensive in situ measurement systems) can be overcome by satellite data. Satellites provide coverage of ocean and land around the world. This can be used to understand time series and trends, and to study local phenomena. One of the satellites looking at sea surface salinity (SSS) is the European Space Agency's (ESA) SMOS mission (Soil Moisture Ocean Salinity), using a Microwave Imaging Radiometer with Aperture Synthesis (MIRAS) from 2010, \cite{esa}. Its resolution is $35\,$km. NASA's Aquarius mission focuses also in the extraction of SSS from space, with a spatial resolution of $150\,$km, but only covered 2011-2015, \cite{nasa}.\\

The MODIS--Aqua mission (Moderate Resolution Imaging Spectroradiometer), \cite{modis}, covers the Earth's surface every 2 days, with information in 36 spectral bands. With operations starting in 2000, the MODIS band information is useful to study aerosols, ocean colour (the surface of the ocean changes colour depending on the chemicals and particles floating in the water), phytoplankton and biochemical properties, water vapour and sea surface temperature (SST). Resolutions of $250\,$m, $500\,$m and $1000\,$m (for ocean reflectance) are provided. SeaWiFS (Sea--Viewing Wide Field--of--View Sensor) \cite{seawifs} has been used to extract SSS from ocean colour. SeaWiFS was designed to collect ocean biological data from 1997 to 2010, focusing on chlorophyll. Information is recorded in 8 bands, at resolutions ranging from $1100\,$m to $4500\,$m. \\

Recent efforts are focusing on techniques to make more accurate predictions of ocean salinity and temperature. The use of artificial intelligence to estimate oceanographic data is an example. \cite{aparna2018} used a neural network to predict sea surface temperature based on measurements of temperature the day before. \cite{aparna2018}. Information for 2 years is predicted with errors around $0.5^{\circ}$C. \cite{garciagorriz2007} predicts sea surface temperature in the Mediterranean using wind data, sea level pressure, dew point temperature, air temperature, and total cloud cover as inputs to the neural network. Predictions on seasonal and interannual SST variability are provided, and these are used to reconstruct incomplete SST satellite tiles (i.e. images). \cite{patil2016} combines numerical estimations from the Indian National Centre for Ocean Information Services (INCOIS) with NOAA's AVHRR SST data to produce temperature in the Indian Ocean by means of a wavelet neural network. Daily, weekly, and monthly temperature values are provided.\\

\begin{figure}[h]
\centering
\begin{tabular}{ll}
	\includegraphics[width=0.85\textwidth]{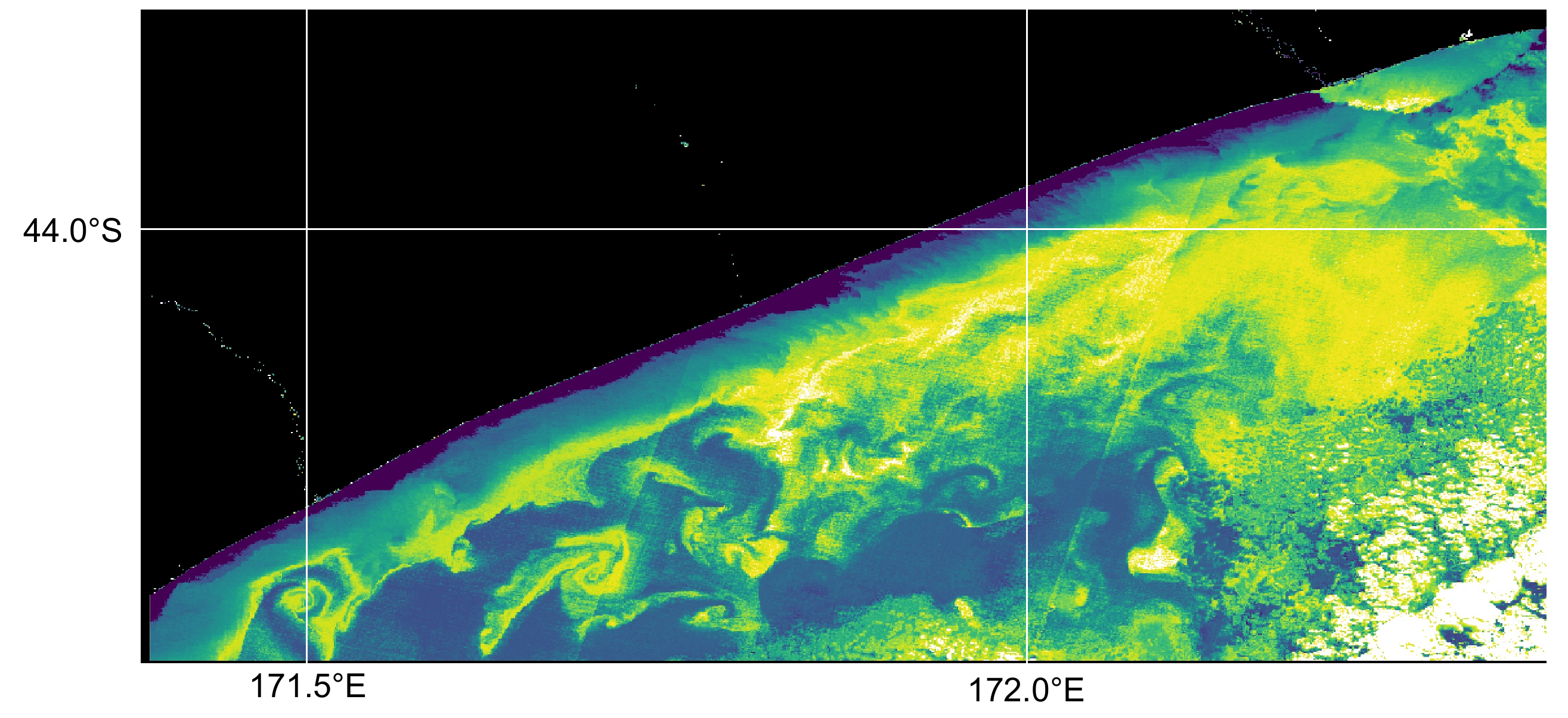}
	\includegraphics[width=0.13\textwidth]{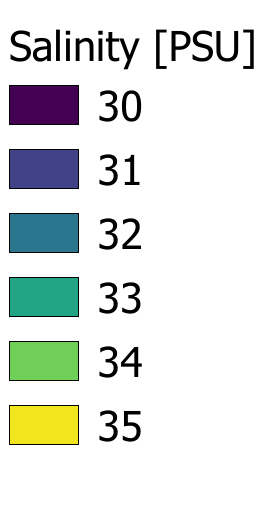}\\\\
\end{tabular}
\caption{Sea surface salinity at Canterbury Bight (New Zealand), $100\,$m resolution. Results using Sentinel--2 Level1C imagery. Land is depicted in black, clouds in white. Extracted from \cite{medina2020}.}
\label{fig_canterbury}
\end{figure}

The amount of research in the area of SSS is less extensive than SST.  \cite{marghany2010} examines the ability of different algorithms to retrieve SSS from MODIS data, comparing with physical measurements from the South China Sea. Other studies use neural networks to predict SSS. \cite{geiger2013} uses salinity data from vessels matched to MODIS--Aqua imagery to train a neural network algorithm that predicts salinity in the Atlantic ocean, with special attention paid to estuaries. \cite{chen2017} presents a neural network that predicts SSS in the Gulf of Mexico. Ocean colour from MODIS--Aqua and SeaWiFS is used, with a resolution of $1000\,$m. Other studies present relevant approaches to obtaining salinity: \cite{olmedo2018} shows a 6--year study on SSS distributions in the Mediterranean from SMOS imagery, limiting errors from previous research. The work is based on Data Interpolating Empirical Orthogonal Functions (DINEOF), and multifractal fusion. Progressing towards higher resolution products, the research published by \cite{medina2019, medina2020} present a methodology to obtain SSS and SST at high--resolution ($100$m) in coastal areas. A neural network is trained for this purpose, using in situ data from Copernicus Marine, and Sentinel-2 data. Determination coefficients about $99\%$ and most common errors about $0.2$PSU (Practical Salinity Unit), \cite{medina2020}, and $0.4^{\circ}C$, \cite{medina2019}, show the potential for the use of machine learning in combination with satellite data to progress in our understanding of physical processes in the ocean, see Figure \ref{fig_canterbury}. Mixing patterns are clearly visible in salinity derived from multispectral sources.\\

\subsubsection{Fish and fisheries}

Fisheries make critical contributions to socio-economic development, food security, nutrition and trade. Despite the signiﬁcant contributions that ﬁsheries provide, they are rarely included in national development policy, due to problems with valuation and access to data, specifically relating to small-scale fisheries. Therefore, a gap in the area of fish and fisheries is the accurate mapping of fishing effort and catches throughout spatial and temporal scales (both annual and seasonal). Fishing vessels are now equipped with a satellite--based monitoring system (VMS), however this is not a requirement for vessels smaller than 12 m. Monitoring the fishing activity and effort through the VMS or automatic identification system (AIS) is key for management of marine activities and contribution to establishing fishing grounds. This type of data would be critical in terms of monitoring fishing activity adjacent to developments, such as OREs, to aid with baseline characterisation and monitor changes in vessel behaviour around OREs to address fisheries displacement issues and uncertainty. Describing fisheries displacement could also be done by utilising top predators’ satellite tags (e.g. birds and mammals) to monitor fish school behaviour within and outside wind farm sites. Previous work \cite{russell2014marine} has shown that seals utilise wind farm seabed structures for foraging. If fitted with tags to examine foraging behaviour, the reef effect on development/recovery time of fish presence around wind farms could also be explored. The application of VMS or AIS data would be critical to aid fisheries monitoring, recruitment to the fishery, stock assessment and movement. In addition, improved understanding of the fishing effort throughout spatial and temporal scales could be utilized in the EIA process to improve assessments related to impacts on commercial fisheries and contribute to the degree of co-existence between commercial fisheries and wind farms.\\

\subsubsection{Mammals and birds}

Changes in animal behaviour, following anthropogenic disturbance, could pose a conservation threat to the population, if the individuals fail to survive, breed or grow. However, methods to understand and predict population-level consequences of such changes are lacking. Individual effects, may be direct and acute, however population-level effects, could lead to changes in abundance and distribution as a result of ORE deployment. However, these effects are much harder to assess on a larger spatial scale and other factors need to be taken into consideration, e.g. quality of alternative habitat, prey availability. In addition, there is a lot of uncertainty relating to collision risk and entanglement research and how local-scale animal behaviour and space use around OREs might change. Therefore, the knowledge gap relates to the local scale behavioural responses and movement of marine animals around OREs, what is the abundance and distribution of the animals in habitats suitable for OREs development, what are the individual consequences of such disturbance and how these individual effects translate into population-level changes at larger spatial scales. \\

Satellite tags could be utilized to gather baseline information on marine animals in suitable areas for ORE development. Such information should ideally be applied across species, seasons and years to better understand animal movement and risk of collision to better inform population models and minimize the potential impact of these activities on both individual and population level. Direct measurements of the presence of birds and marine mammals from satellite imagery are limited to the pixel resolution of available products, whose maximum resolution is hundreds of meters for free data and down to 40 cm for commercial products. It is evident that small and elusive marine mammals, such as seals or dolphins, are difficult to sight using low resolution satellite data, leading most of the research to use unmanned aerial vehicle (UAV) for their monitoring. Moreover, EIAs for top-predators require repeated and standardized measurements that might demand a large amount of effort to define and apply a specific field sampling design. A recent study has demonstrated that it is possible to manually detect four different species of whales from very high resolution satellite imagery (Figure 17C, \cite{cubaynes2019whales}).
Indirect information about the presence of the species can be derived from ecological modelling of the suitable habitat use. Physical and biological variables from remote sensors could be used as proxies to better understand foraging behaviour and habitat preferences of animals and how these vary with weather conditions and climate change. Capturing spatial and temporal variation is key to aid EIA processes and decision-making.\\

\subsubsection{Habitat change}

It is critical to understand how species use their habitat and feeding grounds, what is the home range before, during and after ORE developments. The habitats in question are those that are likely impacted by ORE instalments, such as sand and gravels, stony reefs, mud, kelp and rocky reefs. The loss of habitat might be operation-specific (e.g. initial building phase of offshore wind farms) and will depend on the size of the ORE being installed. A knock-on effect is also likely to happen on habitats outside the building area. Offshore wind farms, in particular, might trigger the formation/development of new habitats, which might be ecologically crucial. In addition, decommissioning might cause habitat loss and changes in the associated species communities, consequently leading to ecosystem-scale effects. Therefore, the choice of suitable habitat for ORE development is critical and monitoring the habitat (along with species communities) before, during and after the ORE development is recommended. Ideally, here the EIA should be utilized at an ecosystem level, thus considering species interactions with each other and their habitat.ORE devices rely on different physical backgrounds, which are spatially and temporally heterogeneous within the marine environment, characterizing patches of habitats that sustain communities of interacting species. For example, tidal devices could cause changes to benthic habitat due to changes in flow and friction around the device, alteration of sediment transport and scouring, but it is uncertain how such localised hydrological conditions will affect species and their habitat. \\

\begin{figure}[h!]
\centering
\includegraphics[width=0.9\textwidth]{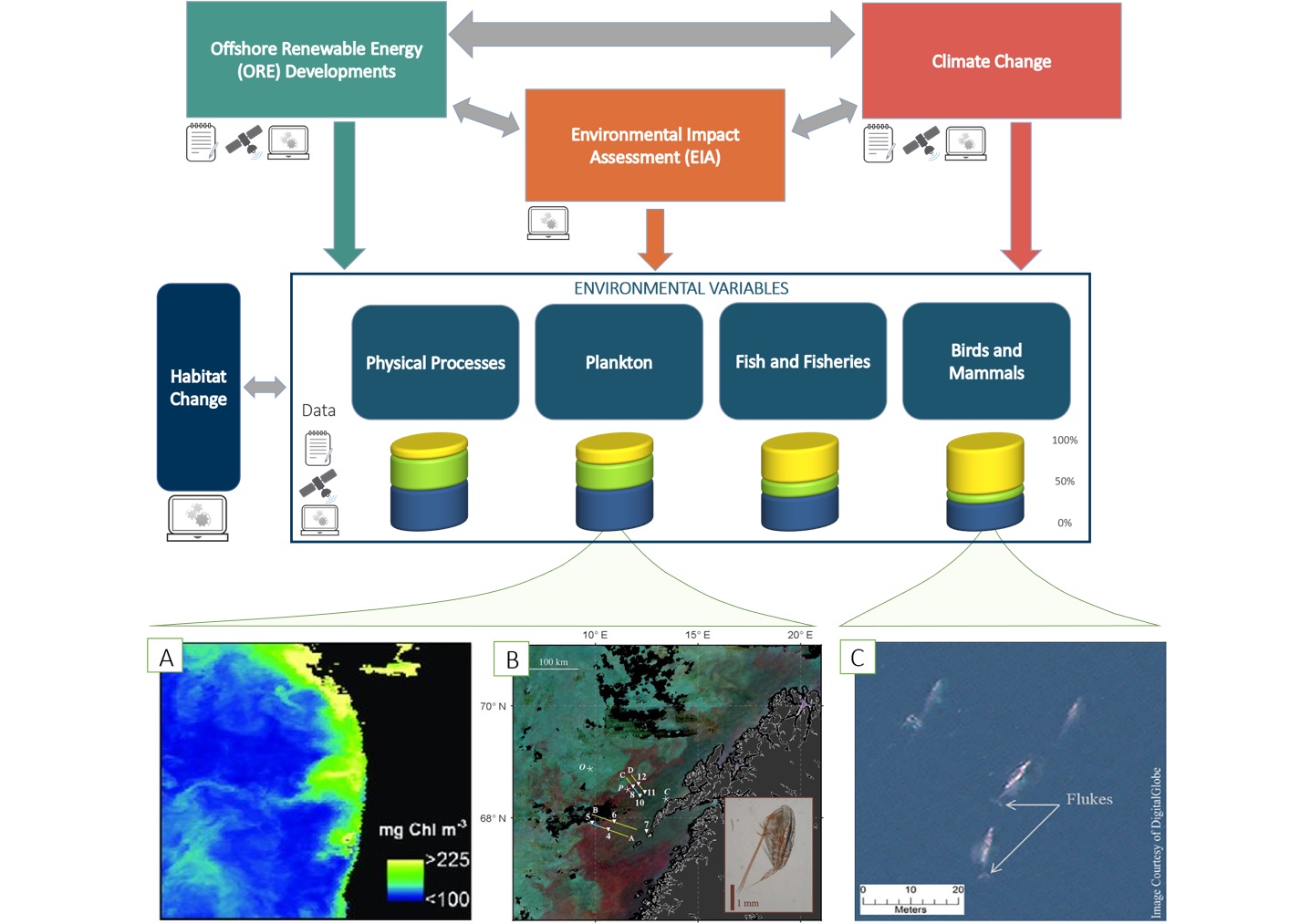}
\caption{The top image represents key environmental variables and their interactions with ORE, EIA and climate change, that have been discussed throughout section \ref{sec:env}. The arrows in grey encapsulate some knowledge gaps in the extent of the interaction that have been addressed throughout \ref{sec:env}. The logos (``notebook", ``satellite" and ``laptop") are used to represent the relevant proportion of data assimilation method to each variable and/or component: \textit{in situ} sampling, satellite imagery or unmanned aerial vehicle and computer modelling, respectively. The bars represent the contribution of each data assimilation method to the environmental variables. The bottom image is an example of satellite images showing (A) different spatial scales of patchiness of chlorophyll a concentration [mg Chl m$^{-3}$] off the coast of Washington-Oregon at 4km resolution detected by MODIS Aqua (\cite{tweddle2018should}, under licence CCBY-4.0), (B) large surface aggregation of zooplankton (\textit{Calanus} spp.) off the coast of northern Norway detected by VIIRS RGB and processed by NEODAAS (\cite{basedow2019remote}, under licence CCBY-4.0), (C) WorldView-3 satellite imagery of 4 gray whales in Laguna San Ignacio (Mexico) (\cite{cubaynes2019whales}, under licence CCBY-4.0).}
\end{figure}

Monitoring habitats and species with the aid of remote sensing and satellite data could be essential to examine changes in the benthic, physical habitat as well as to monitor changes in phenology, distribution or migratory behaviour (through the use of satellite tags). Currently, satellite ocean colour measurements provide records of phytoplankton pigment and carbon concentration from the pelagic global ocean with a spatial resolution of about 1km \cite{muller2018satellite}. Such records are essential in terms of assessing the effects of natural and anthropogenic changes on marine habitats. However, limitations still exist with assessing habitat change over spatial and temporal scales, relevant to human activity. Information gathered by satellites should be used to define and assess these habitats,employing topographic (bathymetry and slope) with environmental variables (sea surface temperature, sea surface colour, sea level change, tidal phases, currents and waves). Such information should be utilized to aid understanding of habitat loss, the creation of new habitats and at what spatial scale (local vs larger scale) population level impacts might arise. Seabed habitat data exist from databases such as EMODNET, \cite{emodnet}, but these are limited to large scale habitat types in offshore waters. Habitat loss, changes and displacement are inevitable consequences of physical, topographic and biological alterations, which still need more considerations during EIA of ORE projects.\\

\section{Conclusions}

This paper presents a holistic view of the current interactions between satellite and ORE sectors, and future needs and opportunities. The paper covers offshore wind, tidal and wave energy technologies, starting with a brief introduction to their history, current developments, most relevant challenges, and future of the sector. Satellite observations for ocean applications are discussed, covering  instrumentation, performance, and data processing and integration. Past, present and future satellite missions are discussed. Both ORE and satellite sectors come together in the final section of the paper, which focuses on innovation opportunities and synergies between them. Uncertainties in standard measurement techniques and the improvements that satellite observations bring are also covered in this work. Wind, wave, and tidal forecasting potential from satellite data are discussed, as well as environmental monitoring capabilities. Applications for Environmental impact assessment of offshore renewables is discussed at length, as satellite observations provide the potential for a more sustainable ORE sector. Measurement of ocean physical properties, as well as fisheries, mammals and birds, and habitats are discussed at the end of the paper, providing a full coverage of all the areas where satellite data can make produce benefits for marine energy.\\ 

The greatest opportunities in the near-term are related to the construction and operation of offshore wind farms. Uncertain site conditions, such as wind and wave climatology, are a factor in design and operational strategy, and short--term forecasts are essential in day-to-day maintenance and energy market participation. Due to the expense and impracticality of in situ measurements, the spatial coverage and resolution offered by satellite data has large potential benefits. Use of existing satellite data to infer wave heights close to individual wind turbines has potential application here, and future missions that measure boundary layer wind profiles will be similarly valuable. In both cases there is a need for long--term monitoring to characterise site conditions and continuous real--time observation for operations and short-term forecasting. \\

While there are products already available that can be used to assess the ORE sector at different stages, and the potential that satellite data is promising, the main challenges that the synergy between satellite and ORE sectors present is the lack of interaction between these, as well as limited knowledge transfer. Satellite data can provide an extra level of information to complement numerical models and in situ data with global coverage, and in many cases, open source. However, the standard techniques used in the ORE sector are well-established and the authors have the opinion that the introduction of new techniques might be difficult, particularly in less developed areas of ORE, such as tidal and wave energy. Through the development of this article, multiple organisations in academia, industry and government in both ORE and satellite sectors came together to set a baseline and common background to promote a better interaction between these sectors, given the multiple interaction points identified in this work. \\

Future steps derived from this work include the promotion and establishment of professional networks to improve communications and collaboration opportunities between satellite and ORE actors. This paper and the project associated have started this type of activities, but the authors would like to encourage further conversations at different levels of development to push for a stronger link between the satellite and ORE sectors. Further work on uncertainty in satellite measurements is needed as well in order to understand how uncertainty propagates into derived products of interest for the ORE sector. This links to data quality assessment and standardisation. The ultimate goal would be to assess the effect of the inclusion of satellite products at different stages of ORE development (e.g. design, construction, operations and maintenance, decommissioning) in reducing the cost of energy. This step is critical to strengthen the position of satellite products in the ORE sector and ensuring the continuity of the application of these methods in future developments. This would also serve as proof of concept for the inclusion of satellite--derived data into international standards for the development of offshore renewables.


\section*{Acknowledgments}

\noindent The authors would like to acknowledge Supergen ORE Hub, grant EP/S000747/1, for the Flexible Funding provided for the SCORE project. G.S. Payne would like to acknowledge the funding from the West Atlantic Marine Energy Community (WEAMEC) of the ``Pays de la Loire'' French Region through the RC+ project. A. Angeloudis would like to acknowledge the NERC Industrial Innovation fellowship grant NE/R013209/2. 

\section*{Note}
\noindent All authors have contributed equally to the paper. The order of authors is the following: the University of Edinburgh--Strathclyde University SCORE team is listed first, and other contributors are then listed in alphabetical order.

\bibliography{mybibfile}

\end{document}